\documentclass[useAMS,usenatbib,times,letter,amssymb]{mn2e}
\usepackage{epsfig,times,amssymb,amsmath,color,verbatim}
\usepackage{xspace,graphicx} % Include figure files

\def\ltsima{$\; \buildrel < \over \sim \;$}
\def\simlt{\lower.5ex\hbox{\ltsima}}
\def\gtsima{$\; \buildrel > \over \sim \;$}
\def\simgt{\lower.5ex\hbox{\gtsima}}
\def\[{\begin{equation}}
\def\]{\end{equation}}

\newcommand{\red}[1]{{\color{black}{#1}}}

\newcommand{\blue}[1]{{\color{black}{#1}}}

\def\m@th{\mathsurround=0pt }
\def\eqalign#1{\null\,\vcenter{\openup1\jot \m@th
\ialign{\strut\hfil$\displaystyle{##}$&$\displaystyle{{}##}$\hfil
\crcr#1\crcr}}\,}
 
 \title[CFHTLenS:  Testing the Laws of Gravity]{CFHTLenS:  Testing the Laws of Gravity with Tomographic Weak Lensing and Redshift Space Distortions}
\author[F. Simpson et al.]{Fergus Simpson$^{1}$\thanks{frgs@roe.ac.uk},
Catherine Heymans$^{1}$,
David Parkinson$^{2}$,
Chris Blake$^{3}$, 
\newauthor Martin Kilbinger$^{4,5,6}$, Jonathan Benjamin$^7$, Thomas Erben$^8$, Hendrik Hildebrandt$^{7,8}$,
\newauthor   Henk Hoekstra$^{9,10}$, Thomas D. Kitching$^1$, Yannick Mellier$^{11}$, Lance Miller$^{12}$, 
\newauthor  Ludovic Van Waerbeke$^7$, Jean Coupon$^{13}$, Liping Fu$^{14}$, Joachim Harnois-D\'{e}raps$^{15, 16}$,
\newauthor   Michael J. Hudson$^{17,18}$, Konrad Kuijken$^{9}$, Barnaby Rowe$^{19,20}$, Tim Schrabback$^{8,9,21}$,
\newauthor  Elisabetta Semboloni$^{9}$, Sanaz Vafaei$^{7}$, Malin Velander$^{12,9}.$\\
\\
$^1$Scottish Universities Physics Alliance, Institute for Astronomy, University of Edinburgh, Royal Observatory, Blackford Hill, Edinburgh, EH9 3HJ, UK. \\ 
$^2$School of Mathematics and Physics, University of Queensland, Brisbane, QLD 4072, Australia\\
$^3$Centre for Astrophysics \& Supercomputing, Swinburne University of Technology, P.O. Box 218, Hawthorn, VIC 3122, Australia\\
$^4$CEA Saclay, Service d'Astrophysique (SAp), Orme des Merisiers, B\^at 709, F-91191 Gif-sur-Yvette, France.\\
$^5$Excellence Cluster Universe, Boltzmannstr. 2, D-85748 Garching, Germany.\\
$^6$Universit\"ats-Sternwarte, Ludwig-Maximillians-Universit\"at M\"unchen, Scheinerstr.~1, 81679 M\"unchen, Germany.\\
$^7$Department of Physics and Astronomy, University of British Columbia, 6224 Agricultural Road, Vancouver, V6T 1Z1, BC, Canada.\\  
$^8$Argelander Institute for Astronomy, University of Bonn, Auf dem H{\"u}gel 71, 53121 Bonn, Germany.\\
$^9$Leiden Observatory, Leiden University, Niels Bohrweg 2, 2333 CA Leiden, The Netherlands.\\
$^{10}$Department of Physics and Astronomy, University of Victoria, Victoria, BC V8P 5C2, Canada.\\
$^{11}$Institut d'Astrophysique de Paris, Universit\'e Pierre et Marie Curie - Paris 6, 98 bis Boulevard Arago, F-75014 Paris, France.\\
$^{12}$Department of Physics, Oxford University, Keble Road, Oxford OX1 3RH, UK.\\ 
$^{13}$Institute of Astronomy and Astrophysics, Academia Sinica, P.O. Box 23-141, Taipei 10617, Taiwan.\\
$^{14}$Key Lab for Astrophysics, Shanghai Normal University, 100 Guilin Road, 200234, Shanghai, China. \\
$^{15}$Canadian Institute for Theoretical Astrophysics, University of Toronto, M5S 3H8, Ontario, Canada.\\
$^{16}$Department of Physics, University of Toronto, M5S 1A7, Ontario, Canada.\\
$^{17}$Department of Physics and Astronomy, University of Waterloo, Waterloo, ON, N2L 3G1, Canada.\\
$^{18}$Perimeter Institute for Theoretical Physics, 31 Caroline Street N, Waterloo, ON, N2L 1Y5, Canada.\\
$^{19}$Department of Physics and Astronomy, University College London, Gower Street, London WC1E 6BT, UK.\\
$^{20}$California Institute of Technology, 1200 E California Boulevard, Pasadena CA 91125, USA.\\
$^{21}$Kavli Institute for Particle Astrophysics and Cosmology, Stanford University, 382 Via Pueblo Mall, Stanford, CA 94305-4060, USA.\\
}

\newcommand{\be}{\begin{equation}}  \newcommand{\ee}{\end{equation}}
  
\newcommand{\ba}{\begin{eqnarray}}\newcommand{\ea}{\end{eqnarray}}
\newcommand{\bm}[1]{\mbox{\boldmath{$#1$}}}   %this is bold italic for MNRAS

\renewcommand{\vec}[1]{{\bmath{#1}}}

\def\gs{\mathrel{\raise1.16pt\hbox{$>$}\kern-7.0pt %
\lower3.06pt\hbox{{$\scriptstyle \sim$}}}}         %
\def\ls{\mathrel{\raise1.16pt\hbox{$<$}\kern-7.0pt %
\lower3.06pt\hbox{{$\scriptstyle \sim$}}}}         %

\newcommand{\ud}{\mathrm{d}}

\newcommand{\om}{\Omega_m}
\newcommand{\ov}{\Omega_\Lambda}
\newcommand{\omk}{\Omega_{\rm{K}}}

\newcommand{\lcdm}{$\Lambda$CDM\xspace}
\newcommand{\wcdm}{$w$CDM\xspace}
\newcommand{\hmpc}{ \, h \rm{ Mpc}^{-1}}
\newcommand{\hinvmpc}{ \, h^{-1} \rm{ Mpc}}

\newcommand{\rhobar}{\bar{\rho}}
\newcommand{\musigma}{($\mu , \Sigma$)\xspace}
\newcommand{\nodata}{~$\cdots$~}
\newcommand{\Pshear}{P_\kappa^{i,j}}
\newcommand{\PsiGR}{{\Psi_{\rm{GR}}}}
\newcommand{\PhiGR}{{\Phi_{\rm{GR}}}}
\newcommand{\DV}{D_{\rm{V}}}
\newcommand{\DA}{D_{\rm{A}}}
\newcommand{\rs}{{r_{\rm{s}}}}
\newcommand{\EG}{{E_{\rm{G}}}}

\begin{document}

\maketitle

\begin{abstract}
Dark energy may be the first sign of new fundamental physics in the Universe, taking either a physical form or revealing a correction to Einsteinian gravity.  Weak gravitational lensing
and galaxy peculiar velocities provide complementary probes of General Relativity, and in combination allow us to
test modified theories of gravity in a unique way.  We perform such an analysis
by combining measurements of cosmic shear tomography from the Canada-France Hawaii Telescope Lensing Survey (CFHTLenS) with the growth of structure from the WiggleZ
Dark Energy Survey and the Six-degree-Field Galaxy Survey (6dFGS),
producing the strongest existing \red{joint} constraints on the metric potentials that
describe general theories of gravity. \blue{For scale-independent modifications to the metric potentials which evolve linearly with the effective dark energy density, we find present-day cosmological deviations in the Newtonian potential and curvature potential from the prediction of General Relativity to be $\Delta \Psi/\Psi = 0.05 \pm 0.25$ and  $\Delta \Phi/\Phi = -0.05 \pm 0.3$  respectively ($68$ per cent CL).}

%For our parameterisation in which modifications to the metric potentials scale with effective dark energy density, we find present-day cosmological deviations in the Newtonian potential and curvature potential from the prediction of General Relativity to be $\Delta \Psi/\Psi = 0.05 \pm 0.25$ and  $\Delta \Phi/\Phi = -0.05 \pm 0.3$  respectively ($68$ per cent CL).
 \end{abstract}


\clearpage

\section{Introduction}
\label{Introduction}

Einstein's original formulation of General Relativity (GR) provides a remarkably precise prescription for the motions of particles in our solar system \citep{Einstein1916}.  It has survived close to a century of experimental scrutiny, beginning with the precession of Mercury's orbit and the gravitational deflection of starlight, progressing to ground-based tests with atomic clocks. Recent spaceborne tests now confirm certain predictions of GR to better than one part in ten thousand \citep{CassiniBertotti}.  % 0.002% Cassini 
However when we consider the behaviour of gravity on cosmological scales, which are over fourteen orders of magnitude greater than the inter-planetary distances, there is much greater scope for uncertainty. Observational evidence of an accelerating universe has called Einstein's laws into question.  The underlying cause for this ``dark energy" phenomenon \blue{may lie within our current understanding of gravitational physics}, the leading candidate being the cosmological constant. Another possibility is that a new regime of gravitational physics has been exposed. It is this prospect which has led to a recent flurry of activity in constraining the cosmological nature of gravity \citep{Daniel2010, Bean2010, Zhao2010, 2010arXiv1011.2106S,  Reyes2010, 2011A&A...530A..68T,  ZuntzISW, Zhao2011, Hudson2012, Rapetti2012, Samushia2012}.

Much like in the solar system, we are able to study the cosmological trajectories of both relativistic and non-relativistic particles, at least in a statistical sense.  The peculiar motions of galaxies falling towards overdense regions generate an illusory anisotropy in their clustering pattern \citep{1987MNRAS.227....1K}, known as redshift space distortions (RSD). This effect has been observed with heightened precision over the past decade \red{\citep{2001Natur.410..169P, 2003MNRAS.346...78H, 2008Natur.451..541G, BlakeWigglezRSD,Beutler2012MNRAS,BOSSRSD2012} permitting inferences of the rate at which cosmological structure forms.  The peculiar velocities of galaxies can also be observed directly, and the inferred measurements of the growth rate are consistent with those from RSDs \citep[see][and references therein]{Hudson2012}.} 

For gravitational experiments, relativistic test particles are of particular interest since their motions are not only sensitive to the effects of time dilation, but also to the curvature of space. In the classical test of GR, the deflection of starlight was observed during a solar eclipse \citep{1920RSPTA.220..291D}. In our experiment, distant galaxies replace nearby stars as the source of photons, our relativistic test particle. When imaging these distant galaxies, instead of measuring their gravitational deflection angles, which are unknown, we can study their shapes which appear correlated as a result of the gravitational deflection of light \citep{BartSchneid}. 

One important caveat is that we are required to convert angles and redshifts into physical distances when interpreting observations of redshift space distortions and gravitational lensing.  We therefore require additional information on the geometry of the universe. Geometric measurements of dark energy such as supernovae and baryon acoustic oscillations, conventionally used to determine the equation of state for dark energy, $w(z)$, are crucial for this purpose.  However their measurements alone cannot distinguish modified gravity models, which are in general capable of reproducing any given \wcdm expansion history.  The Cosmic Microwave Background (CMB) is also used to further break degeneracies, and provides a high redshift anchor for the amplitude of density perturbations \citep[see e.g.][]{Komatsu2011}.

Gravitational lensing is not unique in its ability to probe both metric potentials. CMB photons gain or lose energy as the potentials evolve, a phenomenon known as the Integrated Sachs Wolfe effect (ISW), leaving a significant anisotropy on the largest angular scales. \blue{In the context of a flat \lcdm background}, the expected signal strength from GR is modest\blue{, while many alternative theories of gravity can readily} generate very large ISW signals, something which is not seen in the data \red{\citep{2009PhRvD..80f3536L,ZuntzISW}.} 

While we currently lack a specific model which can compete with GR in terms of simplicity and physical motivation, there exist broad families of models such as $f(R)$ which provide useful test cases. For a recent review of the full menagerie of models, see \citet{CliftonReview}. In this work we do not aim to constrain the parameter space of a particular family of models, or explicitly perform Bayesian model selection, but instead simply address the question of whether our data appear consistent with the predictions of GR. In addition to the cosmic shear data from the Canada-France-Hawaii Telescope Lensing Survey (CFHTLenS), we make use of redshift space distortions from WiggleZ and the Six-degree-Field Galaxy Survey (6dFGS). 

In \S \ref{sec:param} we outline our choice of parameterisation for departures from GR. Our datasets are summarised in \S  \ref{sec:geometric_data} and \S \ref{sec:gravitational_data}, followed by our methods and main results in \S \ref{sec:methods} and \S \ref{sec:results}. A comparison with alternative parameterisation schemes is presented in \S \ref{sec:alt_params}, and we briefly review the status of theoretical models of modified gravity in \S \ref{sec:theory}, with concluding remarks in \S \ref{sec:conclusions}.

\section{Parameterising the Modification}
\label{sec:param}

\begin{figure}
\includegraphics[width=80mm]{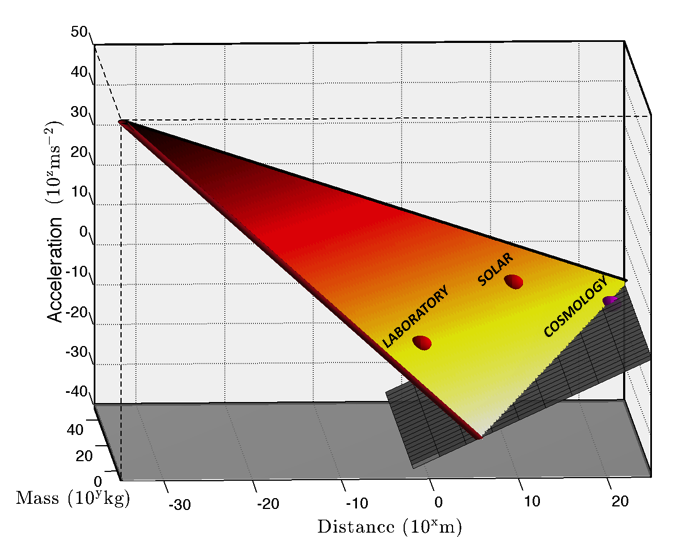}
\caption{ \label{fig:logplot} The shaded triangle represents the vast regime over which the gravitational acceleration of a test particle in the proximity of a mass M is thought to be well described by the Newtonian prescription $a = GMr^{-2}$. While torsion balance (Laboratory) and space-based (Solar System) tests of gravity offer superior accuracy, cosmological methods push the boundaries of our understanding of gravity. For clarity we truncate the mass axis at the Planck mass (red edge),  while the upper bound (black edge) denotes the Schwarzschild radius. The grey slope illustrates the amplitude of the negative contribution arising from dark energy, which scales in proportion to $r$, and surpasses the conventional Newtonian force at cosmological scales.}
\end{figure}

Given the lack of a compelling model to rival GR, we choose to parameterise deviations from GR in a phenomenological manner. In effect, we simply wish to ask: is the strength of gravity the same on cosmological scales as it is here on Earth?  If not, this may modify the motions of both relativistic and non-relativistic particles, but not necessarily in the same manner. Figure \ref{fig:logplot} illustrates why we are seeking a modified gravity signal on cosmological scales; it is the regime in which the classical Newtonian attraction is overwhelmed by the repulsive force associated with dark energy. 

The perturbed Friedmann-Robertson-Walker metric may be expressed in terms of the scale factor $a(t)$, the Newtonian potential $\Psi$, and the curvature potential $\Phi$

\[ \label{eq:FRW}
ds^2 = (1+2\Psi)\, dt^2 - a^2(t)\, (1-2\Phi)\, d \vec{x}^2 \, .
\]

\noindent In GR each of the two gravitational potentials  $\Psi$ and $\Phi$, which carry implicit spatial and temporal dependencies, may be determined from the distribution of matter in the Universe. In Fourier space this is given by

\[
k^2 \Phi_{GR} =  - 4 \pi G a^2 \rhobar \delta  \, ,
\]

\noindent where we have introduced the wavenumber $k$, the mean cosmic density $\rhobar$, and the fractional density perturbation $\delta \equiv \rho/\rhobar - 1$. In the absence of anisotropic stress  the two potentials are identical, $\Psi (x, t) = \Phi (x, t)$. Relativistic particles collect equal contributions from the two potentials, since they traverse equal quantities of space and time, such that

\[
k^2 \left(\PsiGR + \PhiGR \right) = - 8 \pi G a^2 \rhobar \delta  \, .
\]

\noindent We wish to investigate deviations from GR with a general parameterisation. Hereafter, each of the two probes, the Newtonian potential $\Psi$ experienced by non-relativistic particles and the lensing potential $\left( \Phi + \Psi \right)$ experienced by relativistic particles, are now modulated by the parameters $\Sigma$ and $\mu$

\[ \label{eq:mod_psi}
\Psi (k, a)  =  \left[1 + \mu (k, a) \right] \PsiGR (k, a)  \, ,
\]

\[ \label{eq:mod_psiphi}
\left[ \Psi (k, a)  + \Phi (k, a)  \right] = \left[1 + \Sigma (k, a)  \right] \left[\PsiGR (k, a)  + \PhiGR (k, a)  \right] \, ,
\]

\noindent where we have adopted notation similar to that of \citet{2008JCAP...04..013A}, as used more recently by \citet{Zhao2010} and \citet{2010arXiv1011.2106S}. This parameterisation has the advantage of separating the modified behaviour of non-relativistic particles, as dictated by $\mu(k,a)$, from modifications to the deflection of light, as given by $\Sigma(k,a)$. By being able to reproduce a wide range of observational outcomes, in terms of the two-point weak lensing correlation functions and the growth of large scale structure, we ensure a broad sensitivity to different types of deviations from GR. There is some flexibility in how we choose to parameterise the scale and time dependence of these two parameters, much like the dark energy equation of state $w(z)$.  Previous works have often \red{chosen a scale independent model with a parameterised time variation} of these functions such that $\Sigma(a) = \Sigma_s a^s $ and $\mu(a) = \mu_s a^s$.  However given that our primary motivation for modified gravity is to seek an explanation of the dark energy phenomenon, we model the time-evolution to scale in proportion with the effective dark energy density implied by the background dynamics, such that

\[ \label{eq:mgparams}
\Sigma(a) = \Sigma_0 \frac{\Omega_\Lambda(a)}{\Omega_\Lambda}   \, , \, \, \, \, \mu(a) = \mu_0 \frac{\Omega_\Lambda(a)}{\Omega_\Lambda}  \, ,
\]

\noindent where we have defined $\Omega_\Lambda \equiv \Omega_\Lambda (a=1)$; this normalisation was chosen such that $\mu_0$ and $\Sigma_0$ reflect the present day values of $\mu(a)$ and $\Sigma(a)$ respectively. Note that for the case of GR, $\Sigma_0 = \mu_0 = 0$. One advantage of this parameterisation is that it permits a trivial mapping between $\mu_0$ and the popular growth index $\gamma$ \citep{1998ApJ...508..483W, 2005PhRvD..72d3529L}. It also \blue{reduces the total number of degrees of freedom, thereby allowing} us to maintain a manageable error budget. \blue{The disadvantage is that we become less sensitive to deviations from GR whose time evolution differs significantly from the adopted model.} The lack of scale dependence in this model may seem questionable given the stringent Solar System tests which must be satisfied.  However the length scales probed by our cosmological datasets are over fourteen orders of magnitude greater, leaving ample opportunity for a transition at scales smaller than those studied here.  Furthermore, any significant departure from scale-independent growth is more readily apparent from studies of the shape of the galaxy power spectrum. 

With our modified potentials  in equations (\ref{eq:mod_psi}, \ref{eq:mod_psiphi}) the growth of linear density perturbations is now given by 

\[ \label{eq:modifiedgrowth}
\delta''(a) + \left(\frac{2}{a} + \frac{\ddot{a}}{\dot{a}^2} \right) \delta'(a) - \frac{3 \om}{2 a^2} \bm{\Big[1 + \mu(a)  \Big]} \delta(a) = 0   \, ,
\]

\noindent while the cosmic shear power spectrum  $\Pshear (\ell)$ correlating redshift bins $i$ and $j$ takes the form

\[ \label{eq:modifiedshear}
\eqalign{
\Pshear (\ell) &= \frac{9}{4} \om^2 \left( \frac{H_0}{c} \right)^4 \int_0^{\infty} \frac{g_i(\chi) g_j(\chi)}{a^2(\chi)} P_\delta \left(\frac{\ell}{f_K(\chi)} , \chi \right)  \cr
& \times \bm{\left[1+\Sigma(\chi) \right]^2}  \ud \chi \, ,
}
\]

\noindent where the terms not present in the equations' conventional form are highlighted in bold.  Here $\chi$ denotes the radial coordinate distance, while $f_K$ is the comoving angular diameter distance. Note also that $P_\kappa (\ell)$ possesses an implicit dependence on $\mu$ via the time-evolution of the $P_\delta$ term. The lensing efficiency $g_i$ is determined by the radial distribution of source galaxies $n_i(\chi)$ and the comoving angular diameter distance $f_K(\chi)$

\[ \label{eq:lensefficiency}
g_i (\chi) = \int_\chi^{\infty} \ud \chi' n_i(\chi') \frac{f_K(\chi'-\chi)}{f_K(\chi')}  \, .
\]

Note that our parameters $\mu_0$ and $\Sigma_0$ are only modulating the gravitational potentials in equation (\ref{eq:FRW}), but not the evolution of the expansion $a(t)$. The standard cosmological model, in the form of a flat \lcdm universe, is known to provide a good description of the cosmic geometry, therefore for most of this paper we work on the assumption of this global expansion, and focus on exploring the less well determined behaviour of the gravitational perturbations. We shall also explore the consequences of adding further degrees of freedom to the expansion history, in the form of an effective dark energy equation of state $w$ and the global curvature $\omk$.

\section{Geometric Data} 
\label{sec:geometric_data}

When considering the prospect of modified gravity, we must relax the assumption that light deflection and structure formation follow the predictions of GR. This significantly exacerbates the task of  constraining all of our unknown cosmological parameters. For example, without knowing how mass bends light on cosmological scales, we cannot hope to use gravitational lensing to measure the geometry of the Universe. We therefore require auxiliary data to assist in measuring the conventional cosmological parameters, those which govern the cosmic expansion history and shape of the matter power spectrum, before we can attempt to tackle the gravitational parameters $\mu_0$ and $\Sigma_0$.

\subsection{Hubble Parameter}

The current rate of cosmological expansion has long been a source of great uncertainty, largely due to the inherent difficulty in performing distance measurements. Considerable improvements have been made in the past decade, and we make use of a recent result obtained from combining observations of local supernovae, Cepheid variables, and the megamaser at the centre of NGC 4258 \citep{2011ApJ...730..119R}. Where stated, we include a Gaussian prior  $H_0 = 73.8 \pm 2.4 \mathrm{\, km \, s^{-1} \, Mpc^{-1}}$ from \citet{2011ApJ...730..119R}.  In de Sitter space this would correspond to the distance between any two comoving points doubling every thirteen billion years.

Since it is common practice to assume GR in most astronomical and cosmological studies, extra care must be taken when adopting quoted results from previous studies. For example, the determination of the distance to water masers is based on the assumption of Keplerian motions. Furthermore the pulsation of Cepheids relies on the star's self-gravitation, therefore unusual gravitational behaviour could lead to a modification of their period-luminosity relation. However there are currently no signs of anomalous behaviour in the data \citep{Bhuv2012}.  

\subsection{Baryon Acoustic Oscillations}
\label{sec:BAO}

%153 Mpc used as fiducial for BOSS
Encoded within the distribution of galaxies is a useful comoving ruler, a modest excess of galaxy pairs at ${\sim}150$ Mpc, relating to the distance sound waves were able to propagate prior to recombination. In Fourier space this feature is referred to as the Baryon Acoustic Oscillations (hereafter BAO). For a given BAO observation, the relevant cosmological information can often be distilled to a single datapoint, that of $\DV (z) / \rs$, where $r_s$ is the sound horizon at the baryon drag epoch, and the volume element  $\DV (z)$ is defined in terms of the angular diameter distance $\DA (z)$ and Hubble parameter $H(z)$

\[
\DV(z) \equiv  \left[(1 + z)^2 \DA^2(z) \frac{cz}{H(z)}\right]^{\frac{1}{3}} \, .
\]

\noindent The most conservative treatment of BAO data would be to treat $r_s$ as an unknown quantity, and eliminate it by taking the ratio of $\DV (z)$ at two separate redshifts. However since the CMB provides us with a good understanding of the conditions in the early Universe, we have the opportunity to determine $r_s$ for a given set of cosmological parameters. For our purposes this quantity is derived using equation (6) from \citet{1998ApJ...496..605E}. 

Where stated, we utilise measurements of $\DV (z)/\rs$ at two different redshifts.  \citet{BOSS2012} present BAO constraints from the Baryon Oscillation Spectroscopic Survey (BOSS), and we also include a measurement from the Sloan Digital Sky Survey (SDSS) Luminous Red Galaxies \citep{Padmanabhan2012}. Both studies present results before and after a reconstruction method has been applied, a technique which increases the prominence of the BAO by partially reversing the effects of gravitational infall. We conservatively adopt the BAO results prior to reconstruction, since we wish to avoid making assumptions relating to the growth of density perturbations.  The change in our results is negligible, however, when the reconstructed values are included. For the BOSS CMASS sample, the `consensus' result \citep{BOSS2012} averages the power spectrum and correlation function analyses, giving
%$r_s / D_v(0.275) = 0.139 \pm 0.0037$ and $D_v(0.35) / D_v(0.2) = 1.736 \pm 0.065$
$\DV / \rs (z=0.57) = 13.67 \pm 0.22$.  The SDSS datapoint is given by $\DV / r_s (z=0.35) = 8.89 \pm 0.31$. We neglect any covariance in the two data points generated by the small overlap between the SDSS and BOSS volumes, which spans less than $9\%$ of the BOSS volume \citep{BOSS2012}.

It should be noted that when combined with WMAP7, recent BAO results suggest values of $H_0$ which are in mild tension with the measurement from \citet{2011ApJ...730..119R}. For example \citet{BOSS2012}  find $H_0 = 68.4 \pm 1.3 \mathrm{\, km \, s^{-1} \, Mpc^{-1}}$, while studies from 6dFGS and SDSS measure $67 \pm 3.2 \mathrm{\, km \, s^{-1} \, Mpc^{-1}}$  and $69.8 \pm 1.2 \mathrm{\, km \, s^{-1} \, Mpc^{-1}}$ respectively \citep{Beutler2011, 2012arXiv1202.0092M}. We present separate results generated using the $H_0$ and BAO data in Section \ref{sec:results}.

\subsection{Cosmic Microwave Background $(\ell > 100)$}

We utilise the temperature $C_\ell^{TT}$ and polarisation $C_\ell^{TE}$ spectra from the WMAP 7-year data \citep{LarsonWMAP7}. This serves two purposes, first it offers strong constraints on a number of cosmological parameters, but it also serves to directly assist in the measurement of the growth of structure.  Our current position in space and time, $z=0$, acts as the benchmark from which all cosmological distance measurements are made. However when mapping the growth of structure, it is the epoch of recombination at $z \simeq 1100$ which effectively acts as our starting point.   The CMB provides a precise measure of the amplitude of density fluctuations in the early Universe. Combining this with a measure of the clustering amplitude at low redshift provides a strong lever for constraining the growth rate between these two epochs. 

Large angular scales of the CMB are sensitive to modified gravity models, and these are treated separately in \S \ref{sec:isw}. The implications for modified gravity have already been studied  \citep[see for example][]{ZuntzISW}, and here our primary focus is to explore the constraints available from lensing. For the range of values considered in this study, $\mu_0$ and $\Sigma_0$ were found to impose negligible change in either $C_\ell^{TT}$ or $C_\ell^{TE}$ beyond the $\ell=100$ multipole (see Section \ref{sec:isw}). Unless otherwise stated we simply truncate the largest angular scales, such that  only $\ell \geq 100$ are included in our analysis.  In doing so we negate the impact of the ISW, and utilise WMAP7 only to assist in measuring the conventional cosmological parameters.  For further details of our treatment of the CMB, see Appendix \ref{sec:appendix}.

\section{Gravitational Data} \label{sec:gravitational_data}

\subsection{Cosmic Shear}

\begin{figure*} 
\includegraphics[width=175mm]{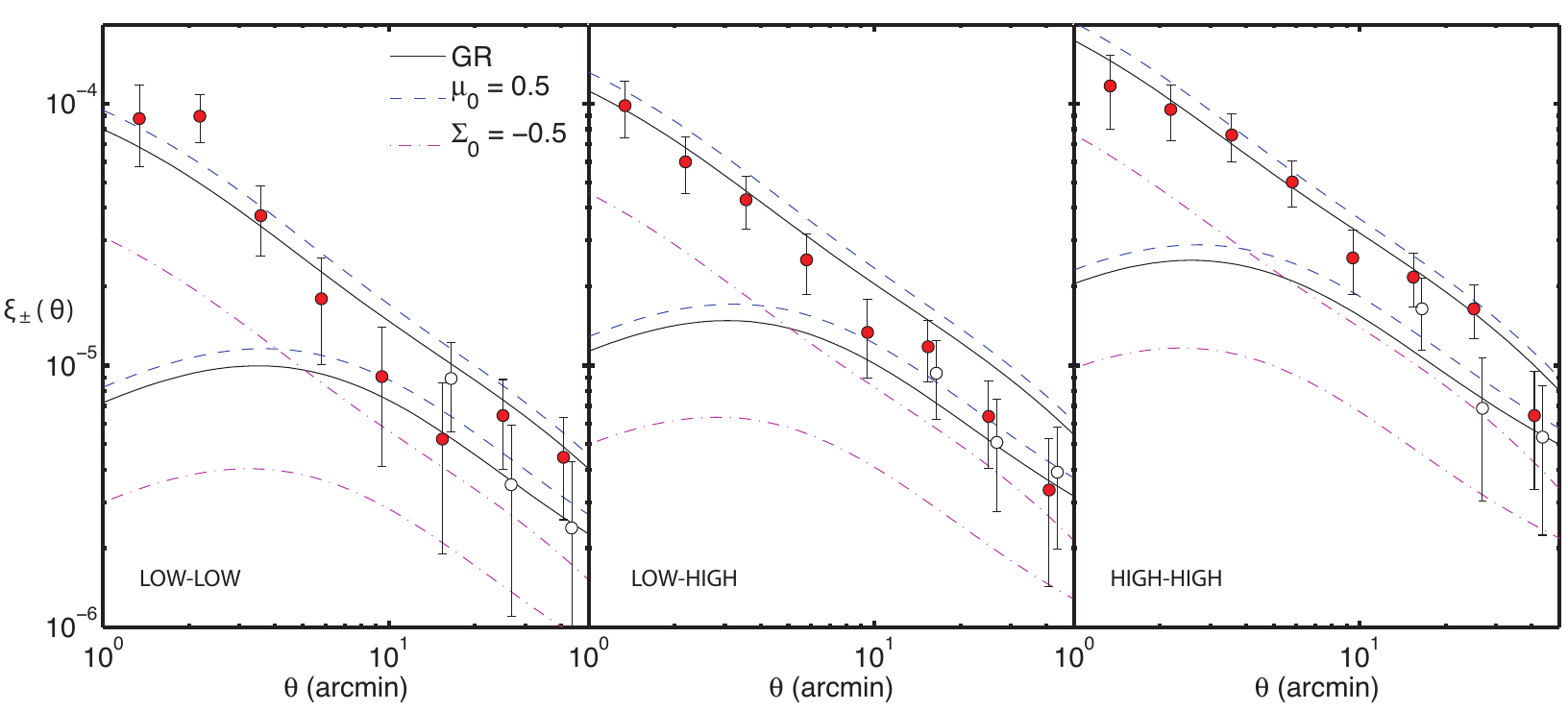}
 \caption{The cosmic shear data from CFHTLenS. \red{Each panel contains the two lensing correlation functions, $\xi_+(\theta)$ (filled) and $\xi_-(\theta)$ (empty) with the $\xi_-(\theta)$ points slightly offset for clarity.  The source galaxies are divided into two redshift bins, a `low' redshift bin with $0.5<z \leq 0.85$ and a `high' redshift bin with $0.85<z \leq 1.3$.  The left, middle and right hand panels display data from the low-low, low-high, and high-high correlations. }In each case, the black lines are the set of theoretical curves corresponding to $\xi_+(\theta)$ (upper) and $\xi_-(\theta)$ (lower) for the best-fitting \lcdm cosmology. The dashed lines correspond to a cosmological model with the same expansion history $H(z)$ as the best-fitting model, but with a boost in the effective gravitational constant ($\mu_0 = 0.5; \Sigma_0 = 0$). The dot-dashed lines demonstrate the signal one expects if the present-day ($z=0$) gravitational potentials are associated with Newtonian gravity ($\mu_0 = 0; \Sigma_0 = -0.5$).\label{fig:cfht_data}}
\end{figure*}

Gravitational lensing by large scale structures in the Universe coherently distort the observed shapes of distant galaxies.  This cosmic shear distortion can be used to directly probe dark matter, and, when analysed in tomographic redshift bins, can also reveal the growth of structure. It also acts as a useful tool for testing the laws of gravity \citep{2001ApJ...560..539W, 2007arXiv0712.1599D}.  We use tomographic cosmic shear data from CFHTLenS, published in \citet{Benjamin2012}.  CFHTLenS is a 154 square degree deep multi-colour survey optimised for weak lensing analyses, where the data were obtained as part of the CFHT Legacy Survey on the 3.6m CFHT.  Weak gravitational lensing is widely recognised as one of the most powerful probes of cosmology, but also one of the most technically challenging, as the weak cosmic shear distortion must be separated from other sources of image distortion induced by the atmosphere, telescope and detector.  The CFHTLenS analysis represents the current state-of-the-art in weak lensing data processing \citep{Erben2009}, weak lensing measurement \citep{Miller2012}, photometric redshift measurement \citep{HH12}, and systematic error analysis \citep{syspaper}.   \citet{Benjamin2012} also use an angular cross-correlation analysis to verify the accuracy of the measured redshift distributions that we use in this analysis.  These results, in combination with the cosmology insensitive joint shear-redshift systematic test presented in \citet{syspaper}, allows us to conclude that the analysis of the survey data is robust within our uncertainties and residual systematics levels and can therefore be used to test the laws of gravity.

\citet{Benjamin2012} analyse two high-redshift bins with $0.5<z \leq 0.85$ and $0.85<z \leq 1.3$, presenting the auto-correlation and cross-correlations of the angular two-point shear correlation functions $\xi_{\pm}^{ij}(\theta)$ over scales $1<\theta<40$ arcmin, \red{as shown in Figure~\ref{fig:cfht_data}}.  The choice to use only two broad high redshift bins in this tomographic analysis was motivated by the desire to mitigate the impact of intrinsic galaxy alignments on the tomographic measurements \citep[see for example][and references therein]{HeymansIA06,BJ11}.  For this combination of high redshift bins, the contamination to the signal is expected to be at the percent level and so we follow \citet{Benjamin2012} by neglecting this source of error in our analysis.  We use a covariance matrix between the resulting correlated data points as estimated from N-body lensing simulations \citep{Clone2012}, discussed and verified in \citet{Kilbinger2012} and \citet{Benjamin2012}.

Theoretical predictions for these correlation functions are generated using the relation

\[
\xi^{i,j}_\pm (\theta) = \frac{1}{2 \pi} \int_0^\infty  P^{i,j}_\kappa (\ell)  \, \rm{J}_{0/4}(\ell \theta) \,  \ell \, \ud \ell \, ,
\]

\noindent where $P^{i,j}_\kappa (\ell)$ is given in equation (\ref{eq:modifiedshear}) and $\rm{J}_{0/4}$ are the zeroth and fourth order Bessel functions.

\red{On the smallest angular scales studied by cosmic shear it is likely that baryonic physics will disturb the density perturbations \citep{2011MNRAS.417.2020S}.  This is also the regime where our prescription for modelling the non-linear correction to the matter power spectrum \blue{(see below)} begins to break down. Due to the very different nature of the zeroth and fourth Bessel functions, the $\xi_{-}(\theta)$ correlation function is sensitive to considerably smaller scales than  $\xi_{+}(\theta)$  for a given $\theta$ value. For the $\xi_-$ statistic we therefore only include the larger angular scales $\theta > 10'$. These data points are presented in Figure \ref{fig:cfht_data}, alongside theoretical predictions from the best-fit \lcdm cosmology and two illustrative modified gravity scenarios. From left to right, the three panels in Figure \ref{fig:cfht_data} relate to the correlations of the low-low, low-high, and high-high redshift bins. A  more detailed justification for the angular cuts and their implications are discussed in \citet{Benjamin2012}; in any case we find no significant change in our results when the small angular scales of $\xi_-(\theta)$ are included due to the high correlation between the points and their relatively low signal-to-noise. We further investigate the impact of truncating smaller length scales in Kitching et al (in prep). }

\subsection{Redshift Space Distortions} \label{sec:WiggleZ}

Doppler distortions of the galaxy clustering pattern provide an independent probe of gravitational physics, by revealing the coherent flows of galaxies which trace the growth of cosmic structure.  The principal effect on large scales is an amplification of apparent galaxy clustering along the line-of-sight, breaking the statistical isotropy.  This anisotropy can be extracted from the 2-point clustering statistics of a galaxy redshift survey, yielding a measurement of $f(z) \sigma_8(z)$.  Here $f$ is the growth rate of structure, expressible in terms of the growth factor $D(a)$ as $f \equiv d {\rm ln} D/d {\rm ln} a$ where the growth factor describes the evolution of the amplitude of a single linear density perturbation $\delta(a) = D(a) \delta(1)$. The function $\sigma_8(z)$ quantifies the amplitude of the linear clustering of matter at a given redshift, and is defined as the r.m.s.\ density variation in spheres of comoving radius $8 \, \hinvmpc$. The degeneracy between $f$ and $\sigma_8$ arises due to the unknown galaxy bias factor $b$: the observables are the anisotropy in the galaxy clustering pattern which depends on $\beta \equiv f/b$, and the galaxy clustering amplitude which depends on $b \sigma_8(z)$.

Measurements of redshift-space distortions from a galaxy redshift survey are degenerate with a second isotropy-breaking mechanism: the Alcock-Paczynski distortion \citep{1979Natur.281..358A}, which arises if the fiducial cosmological model we adopt for converting redshifts and angles to physical distances differs from the true cosmological model \citep{1996MNRAS.282..877B, 1996ApJ...470L...1M, 2003ApJ...598..720S, simpsonp09}.  The amplitude of Alcock-Paczynski anisotropy depends on the quantity 

\[
F(z) = (1+z) \DA (z) H(z)/c \, ,
\]

\noindent in terms of the angular diameter distance $D_A(z)$ and Hubble expansion parameter $H(z)$.  Although measurements of $F$ and $f \sigma_8$ from a galaxy redshift survey are strongly correlated, the angular distortions imprinted in the clustering pattern by these two effects are sufficiently different that the degeneracy can be broken.  

\begin{figure} 
\includegraphics[width=80mm]{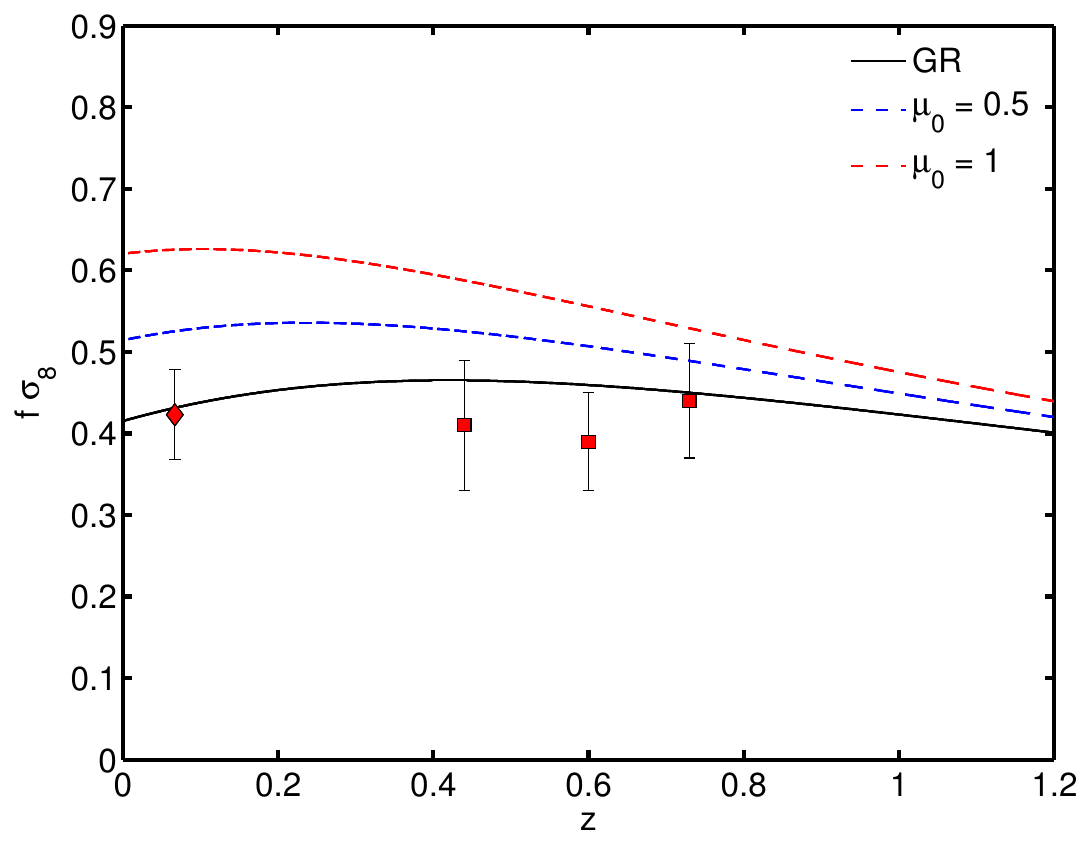}
 \caption{The growth data from WiggleZ \blue{\citep{Blake2012MNRAS}} and 6dFGS \citep{Beutler2012MNRAS}. The black line is the corresponding theoretical curve for the best-fit \lcdm cosmology. The dashed lines correspond to the same cosmology but with a boost in effective gravitational constant $\mu_0$.  \label{fig:fsig8_data}}
\end{figure}

\red{The measurements of ($f \sigma_8$, $F$) that we use in this study are derived from the WiggleZ Dark Energy Survey published in  \citet{Blake2012MNRAS} and from 6dF Galaxy Survey (6dFGS), as presented in \citet{Beutler2012MNRAS}.  Figure~\ref{fig:fsig8_data} compares the measurements of $f \sigma_8$ from WiggleZ (squares) and the 6dFGS (diamond) to the best-fit $\Lambda$CDM cosmology and two example modified gravity scenarios with $\mu_0=0.5$ and $\mu_0=1.0$ (dashed)}.

The WiggleZ Survey \citep{Drinkwater2010} at the 3.9-m Anglo-Australian Telescope has mapped a cosmic volume ${\sim} 1$ Gpc$^3$ over the redshift range $0 < z < 1$. By covering a total of ${\sim}800$ deg$^2$ of sky the WiggleZ survey mapped about 100 times more effective cosmic volume in the $z > 0.5$ Universe than previous galaxy redshift surveys.  Target galaxies were chosen by a joint selection in UV and optical wavebands, using observations by the Galaxy Evolution Explorer satellite (GALEX)  \citep{2005ApJ...619L...1M} matched with ground-based optical imaging from the Sloan Digital Sky Survey \citep{2000AJ....120.1579Y} in the Northern Galactic Cap, and from the Red Sequence Cluster Survey 2 (RCS2) \citep{2011AJ....141...94G} in the Southern Galactic Cap.  A series of magnitude and colour cuts \citep{Drinkwater2010} were used to preferentially select high-redshift star-forming galaxies with bright emission lines, that were then observed using the AAOmega multiobject spectrograph \citep{2006SPIE.6269E..14S} in 1-hr exposures.

\red{The 6dFGS is a near-infrared selected combined redshift and peculiar velocity survey covering nearly the entire southern sky with Galactic latitude $|b| < 10^\circ$.  It was undertaken with the Six-Degree Field (6dF) multi-fibre instrument on the UK Schmidt Telescope from 2001 to 2006 \citep{2004MNRAS.355..747J,2009MNRAS.399..683J}.  The redshift survey covers $125{,}000$ galaxies with $8.75 \le K \le 12.75$ and median redshift $z = 0.052$.}

In this study we use the WiggleZ joint measurements of ($f \sigma_8$, $F$) in three different redshift slices with effective redshifts $z = 0.44$, $0.60$, and $0.73$.  The marginalised measurements of the growth rate in these redshift slices are $f \sigma_8(z) = (0.41 \pm 0.08, 0.39 \pm 0.06, 0.44 \pm 0.07)$.  The marginalised measurements of the Alcock-Paczynski distortion parameter are $F = (0.48 \pm 0.05, 0.65 \pm 0.05, 0.86 \pm 0.07)$.  The correlation coefficients between these pairs of values are $(0.73, 0.74, 0.85)$.  These measurements were derived by fitting the non-linear clustering model provided by \citet{JenningsRSD2011} to the 2D WiggleZ power spectrum over the range of wavenumbers $0 < k < 0.2 \, h$ Mpc$^{-1}$, marginalising over the linear galaxy bias factor and a small-scale velocity dispersion $\sigma_{\rm{v}}$.  \citet{BlakeWigglezRSD} carried out a suite of systematics tests to demonstrate that the derived results did not depend on either the fitting range or the non-linear model of redshift-space distortions adopted; we refer the reader to this paper for further details. 
There is some overlap (${\sim}500$ sq. degrees) between the sky areas surveyed by WiggleZ and the SDSS/BOSS surveys that are used in our results which are stated to include BAO. However, based on the nature of the geometric BAO data being very different to that used for the RSD gravitational measurements we assume that any covariance in these derived results is negligible. In essence, when considering the redshift-space galaxy power spectrum, the location of the acoustic peak bears little connection to the power of the quadrupole.  %Furthermore the analysis in \citet{BOSS2012} marginalised over the shape of the clustering pattern.

We also include the single datapoint $f \sigma_8(z=0.067) = 0.423 \pm 0.055$  \citep{Beutler2012MNRAS} from the 6dFGS. Due to its low redshift, this result benefits from having negligible sensitivity to the Alcock-Paczynski distortion, and a large sensitivity to our modified gravity parameter $\mu_0$, \red{as can be seen in Figure~\ref{fig:fsig8_data}.  The 6dFGS measurement was derived using the non-linear model of Jennings et al.\ (2011), including wide-angle corrections,  fit to the 2D galaxy redshift-space correlation function for transverse separations $r_p > 16 \, h^{-1}$ Mpc.  The growth-rate measurements obtained, marginalized over bias, were consistent with fitting an empirical streaming model to the results \citep[see][for further details]{Beutler2012MNRAS}.} \blue{While the WiggleZ and 6dFGS analyses differ in their utilisation of the power spectrum and correlation function respectively, their approach to modelling non-linear corrections and galaxy bias remains consistent.}

\subsection{Cosmic Microwave Background $(\ell < 100)$}
\label{sec:isw}

\begin{figure} 
\includegraphics[width=80mm]{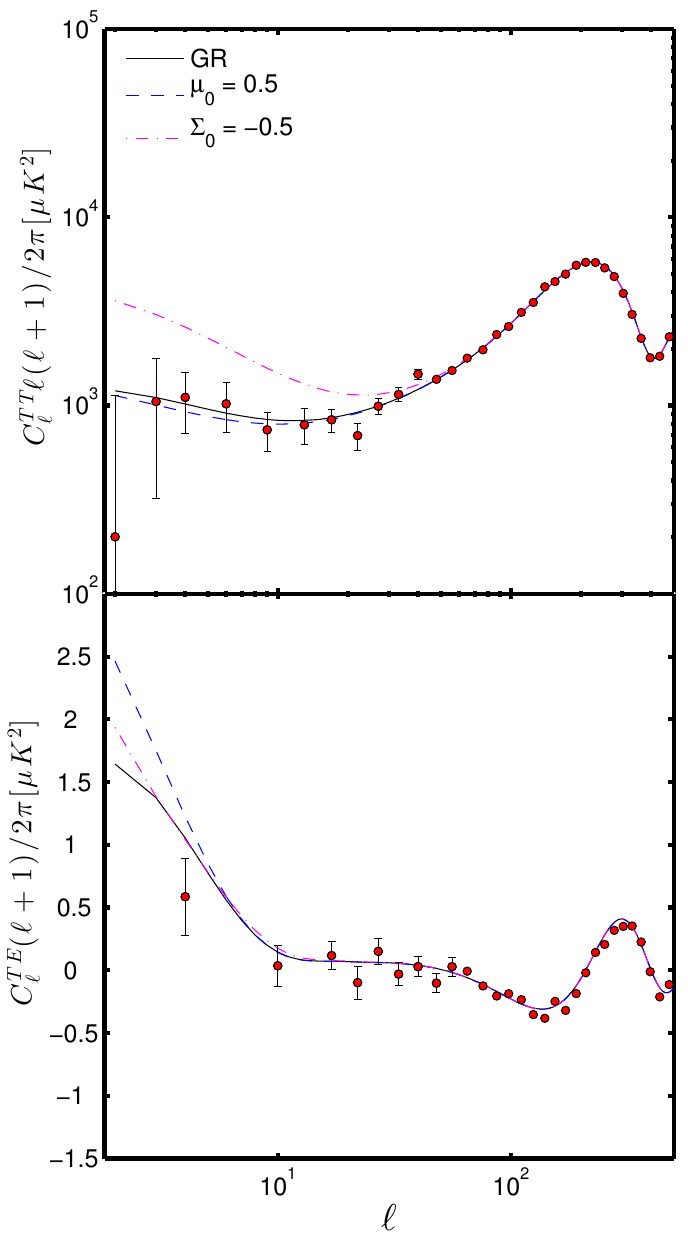}
 \caption{Modifications to the temperature (upper panel) and temperature-polarisation (lower panel) power spectra of the CMB. In each panel the black line is the corresponding theory curves for the best-fitting \lcdm cosmology. The dashed and dash-dot lines represent the same modifications to our modified gravity parameters as seen in Figure \ref{fig:cfht_data}. Superposed  on these theoretical predictions are the binned data points from the $7-$year WMAP observations \citep{LarsonWMAP7}.  \label{fig:cmb_data}}
\end{figure}

When in the presence of a time-dependent gravitational potential, a particle may extract or lose energy from its surroundings. \blue{It is this mechanism which space probes exploit when passing close to planets for a gravitational slingshot.}
For CMB photons, the gravitational source is associated with large scale density perturbations. \blue{In GR, if}
%If
the photon passes through a high density region of the Universe, the dark energy-induced decay of the gravitational potentials $\Phi (x,t)$ and $\Psi (x,t)$ results in a net energy gain, with a blue-shifting effect. Conversely, passing through a void results in a net loss of energy. The resulting temperature perturbation may be expressed by the line integral

\[ \label{eq:ISW}
\frac{\Delta T}{T} = - \int \frac{\ud }{\ud \eta} \left[ \Phi (x, t)   + \Psi (x, t)   \right] \ud \eta \, ,
\]
%$\tau$ is defined by  $\ud \tau \equiv \ud t/a$
\noindent where $\eta$ is the conformal time defined by  $\ud \eta \equiv \ud t/a$, and the integral follows the path of the photon. We can see from equation (\ref{eq:ISW}) that this ISW effect is another probe which, like gravitational lensing, retains sensitivity to both gravitational potentials. However the ISW signal within the temperature anisotropies is only expected to be significant on large angular scales, $\ell < 100$ \red{as illustrated by Figure~\ref{fig:cmb_data} which compares the temperature (upper) and temperature-polarisation (lower panel) GR power spectra of the CMB  (solid) with data points used in this analysis, from the $7-$year WMAP observations \citep{LarsonWMAP7}.  The dashed and dash-dot lines represent the same modifications to our modified gravity parameters as seen in Figure \ref{fig:cfht_data} with $\mu_0 = 0.5; \Sigma_0 = 0$ (dashed) and $\mu_0 = 0; \Sigma_0 = -0.5$ (dot-dashed).  These modified gravity models of the CMB power spectrum deviate from GR only on the largest angular scales.    Observational ISW studies are therefore heavily restricted by cosmic variance, and future experiments cannot be expected to improve upon WMAP, unless we begin to include cross-correlations with large-scale structure. Note that throughout this work, we use the term ISW to refer to the phenomenon which arises in the CMB power spectrum, not the cross-correlation between the CMB and the distribution of galaxies.}

For any metric theory of gravity, density fluctuations on scales beyond the horizon are subject to a stringent consistency relation. Since material cannot be transported at superluminal velocities, superhorizon perturbations are destined to evolve along a path determined by the global expansion \citep{2006ApJ...648..797B}. Therefore the constraints presented which incorporate the ISW should be treated with this in mind, as the ISW involves scales very much larger than those studied by gravitational lensing, a regime where there is a stronger expectation that the values of $\mu_0$ and $\Sigma_0$ are close to zero.

\section{Methods}
\label{sec:methods}

Using a combination of geometric data (presented in \S \ref{sec:geometric_data}) and gravitational data (presented in \S \ref{sec:gravitational_data}), we use a maximum likelihood analysis to constrain our phenomenological modified gravity parameters $\mu_0$ and $\Sigma_0$.

To generate the likelihood values we modified CosmoPMC \citep{KilbingerPMC}, an adaptive importance sampling code, to incorporate the alterations to the growth of structure (equation \ref{eq:modifiedgrowth}) and cosmic shear signal (equation \ref{eq:modifiedshear}).  A customised version of MGCAMB \citep{2011Hojjati, 2000CAMB} was used to evaluate the CMB anisotropies for a given set of cosmological parameters, which are then assessed with the WMAP likelihood code\footnote{http://lambda.gsfc.nasa.gov}. Our analysis spans the parameter set $\{\Omega_m, \Omega_b, h, \tau, n_{\rm{s}}, A_{SZ} , \Delta_{\rm{R}}^2, \mu_0, \Sigma_0 \}$, where $\tau$ is the optical depth, $n_s$ the spectral index, \blue{$A_{SZ}$ the amplitude of the Sunyaev-Zel'dovich template for the CMB \citep{KomatsuASZ, 2007ApJS..170..377S}}, and $\Delta_{\rm{R}}^2$ controls the primordial amplitude of matter perturbations. Where specified we also marginalise over a time-independent effective equation of state for dark energy $w$ and global curvature $\omk$, in addition to the aforementioned set of parameters. \blue{We adopt flat priors throughout, spanning the following parameter ranges:  $\Omega_m \in [0.05; 1]$, $\Omega_b \in [0.01; 0.1]$, $h \in [0.5; 1]$, $\tau  \in [0.02; 0.3]$, $n_{\rm{s}}  \in [0.8; 1.2]$, $A_{SZ} \in [0; 2]$, $\Delta_{\rm{R}}^2 \in [1.8; 3.5] \times 10^{-9}$, $\mu_0 \in [-4.0; 4.0]$, $\Sigma_0 \in [-4.0; 4.0]$, $w \in [-1.5; -0.3]$, and $\omk \in [-0.1; 0.1]$.  In most cases these boundaries extend well beyond the regions of high likelihood, except for $A_{SZ}$ which is not well constrained, however its impact on the other parameters is negligible. In order to generate visualisations of the generated samples} we make use of code adapted from CosmoloGUI\footnote{http://www.sarahbridle.net/cosmologui}. 

In generating the matter power spectrum, the CosmoPMC package makes use of the \citet{1998ApJ...496..605E} transfer function. The growth factor is determined in accordance with the modified equation (\ref{eq:modifiedgrowth}), before adding a nonlinear correction using the prescription of \citet{halofit}. To compensate for cosmologies where $w \neq -1$, a further modulation of the nonlinear power is required, as presented in \citet{2006MNRAS.366..547M}. We make use of the scaling correction provided in \citet{Schrabback2010} to apply this formalism across our full range of parameter space. However it should be noted that our results only have a weak dependence on this correction. The mean values of $\mu_0$, $\Sigma_0$, and $w$ all change by less than $0.02$ when the prescription from \citet{Schrabback2010} is omitted, which is well within our statistical errors. The value of the $\Sigma_0$ parameter has no impact on structure formation, and while we may expect nonzero values of $\mu_0$ to alter the nonlinear correction to the matter power spectrum, we do not attempt to correct for it here. We expect the prescription of \citet{halofit} to remain a reasonably good approximation in the presence of a simple scale-independent modification to the growth of structure such as that induced by $\mu_0$. To first order the correction associated with nonlinear power is simply a function of the ambient linear power at that epoch; how quickly it arrives at this state is of less importance.  In any case, modified behaviour on nonlinear scales would still weaken the fit to the data achievable within the framework of GR.

\section{Results} 
\label{sec:results} 
\begin{figure*} 
\centerline{\includegraphics[width=180mm]{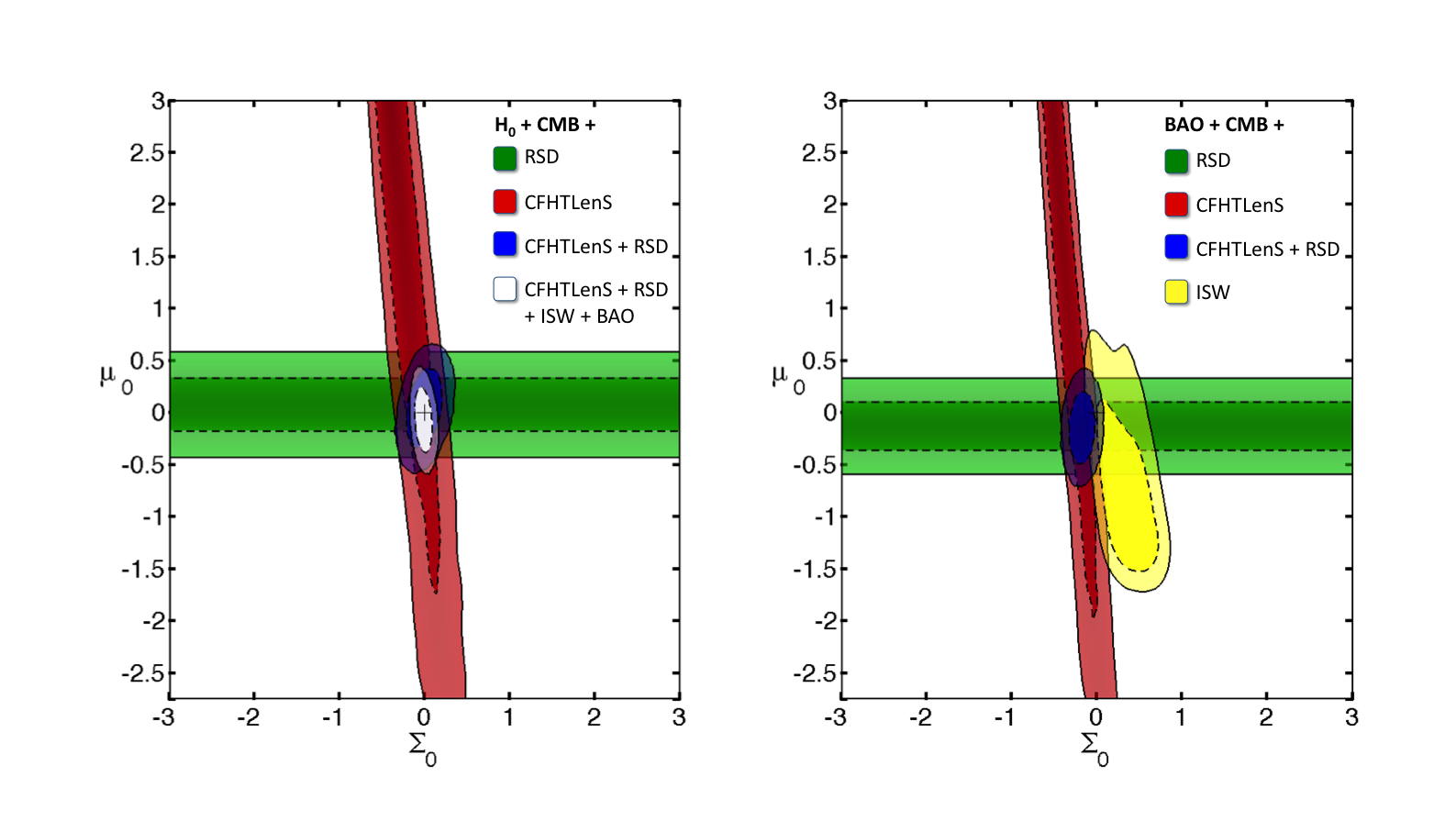}}
% WiggleZ + WMAP7 + $H_0$ + CFHTLenS
 \caption{\emph{Left:} Constraints on the modified gravity parameters in a flat \lcdm background from redshift space distortions (green), weak lensing (red), and combined (blue).  (68 and
95 per cent CL). The dashed and solid contours represent the 68 and 95 per cent conÞdence intervals respectively. Two auxiliary datasets are used here to break degeneracies with the conventional cosmological parameters. These are the small-angle anisotropies from WMAP7 ($\ell \geq 100$), and a prior on $H_0$ from \citep{2011ApJ...730..119R}.  The cross positioned at the origin denotes the prediction of General Relativity. \, \emph{Right:} The red, green and blue contours are the same as the left panel, except the prior on $H_0$ has been replaced by measurements of the Baryon Acoustic Oscillations as detailed in Section \ref{sec:BAO}. The yellow contours signify the constraints derived from the full WMAP7 power spectra, including the large angular scales ($\ell < 100$). \red{The white contours in the left hand panel show the constraints when all data sets are analysed in combination.}  \label{fig1hb}} 
\end{figure*}

\subsection{\lcdm}

We begin with the simple case of a flat \lcdm geometry. In this scenario we are making a strong assumption about the form of the cosmic expansion history $H(z)$, but in doing so make significant gains in the precision of our likelihood contours. Any apparent deviation from GR witnessed at this stage must be interpreted with caution due to the strong assumptions of this relatively simple model.

The left panel of Figure \ref{fig1hb} illustrates the variety of $1$-- and $2$--$\sigma$ confidence intervals for different combinations of datasets. Each of these contours include the auxiliary data of WMAP7 ($\ell \geq 100$) and a prior on  $H_0$. The horizontal (green) contours reflect the pure $\mu_0$ measurement established with the WiggleZ data. The $\Sigma_0$ parameter  is unconstrained since the peculiar motions of galaxies are slow $(v \ll c)$, and are therefore insensitive to the spatial curvature $\Phi(x,t)$. The near-vertical (red) contours correspond to the cosmic shear data from CFHTLenS. Unsurprisingly, this is predominantly a measure of $\Sigma_0$ due to its direct influence on the lensing potential. The mild sensitivity to $\mu_0$ arises from the amplitude of the matter power spectrum entering in equation (\ref{eq:modifiedshear}). A positive value of $\mu_0$  enhances the growth of structure and thus strengthens the cosmic shear signal (for a given primordial amplitude). This can be offset with a negative value of $\Sigma_0$, weakening the deflection of light and leading to the negatively-sloped degeneracy direction seen in the red contours of Figure \ref{fig1hb}. When combined (blue), the near-orthogonal degeneracy directions of WiggleZ and CFHTLenS produce a considerable reduction in the permitted area of parameter space.  The tightest constraints (white) emerge from adding in BAO from BOSS and the ISW large scale anisotropies $(\ell < 100)$ from WMAP7. None of these combinations show any preference for non-zero values of either $\mu_0$ or $\Sigma_0$, suggesting that GR remains a successful model of gravitation on cosmological scales, for both relativistic and non-relativistic particles.

The mean and 1-$\sigma$ errors in the modified gravity parameters are found to be $\mu_0 = 0.05 \pm 0.25$ and $\Sigma_0 = 0.00 \pm 0.14$. The results for other combinations of datasets can be found in Table \ref{tab:results}. 

How sensitive are we to the choice of auxiliary data? The right hand panel of Figure \ref{fig1hb} contains the same datasets and layout as the left hand panel, except here we have replaced the prior on $H_0$ with the BAO data summarised in Section \ref{sec:BAO}. The BAO prefer a higher value of $\om$ (or equivalently, a lower value of $H_0$),  and this does push the lensing data to predict a lower value of $\sigma_8$, which pushes the bulk of the likelihood towards negative values for $\Sigma_0$. This results in constraints of  $\mu_0 = -0.12 \pm 0.23$ and $\Sigma_0 = -0.17 \pm 0.10$. \red{The combination of CFHTLenS and RSD data (blue) can be compared with the constraints derived from the full WMAP7 power spectra, including the ISW, and BAO (yellow).}

\begin{table*}
  \caption{\label{tab:results} Parameter constraints for different combinations of dataset and parameter space. The three different backgrounds we explore are flat \lcdm, flat \wcdm and the non-flat \wcdm which we denote as o\wcdm. \blue{All constraints make use of the small scale anisotropies from WMAP7 ($\ell \geq 100$), while those marked ISW also utilise the larger angular scales.} 
  }
  \begin{tiny}\begin{tabular}{|c|c|c|c|c|c|c|c|c|}
\hline
\textbf{Background}&\textbf{Data (+ \blue{CMB} + $H_0$)}&\textbf{$\mu_0$}&\textbf{$\Sigma_0$}&\textbf{$\Omega_m$}&\textbf{$H_0$}&\textbf{$\sigma_8$}&\textbf{$w$}&\textbf{$\Omega_K$}\\\hline
$\Lambda$CDM&CFHTLenS& 1.2 $\pm$ 2.2 & -0.17 $\pm$ 0.28 & 0.260 $\pm$ 0.024 & 71.3 $\pm$ 2.3 & 0.96 $\pm$ 0.24 &\nodata&\nodata\\\hline
$\Lambda$CDM&RSD& 0.08 $\pm$ 0.25 &\nodata& 0.254 $\pm$ 0.023 & 71.8 $\pm$ 2.2 & 0.801 $\pm$ 0.024 &\nodata&\nodata\\\hline
$\Lambda$CDM&CFHTLenS + RSD& 0.05 $\pm$ 0.25 & 0.00 $\pm$ 0.14 & 0.256 $\pm$ 0.023 & 71.6 $\pm$ 2.2 & 0.804 $\pm$ 0.023 &\nodata&\nodata\\\hline
$w$CDM&CFHTLenS + RSD + BAO& -0.59 $\pm$ 0.34 & -0.19 $\pm$ 0.11 & 0.272 $\pm$ 0.015 & 71.7 $\pm$ 1.8 & 0.820 $\pm$ 0.025 & -1.19 $\pm$ 0.10 &\nodata\\\hline
o$w$CDM&CFHTLenS + RSD + BAO& -0.65 $\pm$ 0.34 & -0.26 $\pm$ 0.12 & 0.289 $\pm$ 0.021 & 71.6 $\pm$ 1.8 & 0.833 $\pm$ 0.028 & -1.16 $\pm$ 0.10 & 0.0096 $\pm$ 0.0087 \\\hline
\textbf{Background}&\textbf{Data (+ \blue{CMB} + BAO)}&\textbf{$\mu_0$}&\textbf{$\Sigma_0$}&\textbf{$\Omega_m$}&\textbf{$H_0$}&\textbf{$\sigma_8$}&\textbf{$w$}&\textbf{$\Omega_K$}\\\hline
$\Lambda$CDM&CFHTLenS& 1.0 $\pm$ 2.2 & -0.31 $\pm$ 0.24 & 0.305 $\pm$ 0.018 & 67.2 $\pm$ 1.4 & 0.95 $\pm$ 0.24 &\nodata&\nodata\\\hline
$\Lambda$CDM&RSD& -0.13 $\pm$ 0.23 &\nodata& 0.300 $\pm$ 0.017 & 67.5 $\pm$ 1.4 & 0.807 $\pm$ 0.023 &\nodata&\nodata\\\hline
$\Lambda$CDM&ISW& -0.62 $\pm$ 0.64 & 0.34 $\pm$ 0.22 & 0.294 $\pm$ 0.015 & 68.2 $\pm$ 1.2 & 0.768 $\pm$ 0.068 &\nodata&\nodata\\\hline
$\Lambda$CDM&CFHTLenS + RSD& -0.12 $\pm$ 0.23 & -0.17 $\pm$ 0.10 & 0.300 $\pm$ 0.017 & 67.6 $\pm$ 1.4 & 0.807 $\pm$ 0.023 &\nodata&\nodata\\\hline
$\Lambda$CDM&CFHTLenS + RSD + ISW + $H_0$& -0.06 $\pm$ 0.21 & -0.010 $\pm$ 0.068 & 0.273 $\pm$ 0.010 & 70.10 $\pm$ 0.97 & 0.803 $\pm$ 0.022 &\nodata&\nodata\\\hline
\end{tabular}
\end{tiny}
\end{table*}

\subsection{\wcdm} \label{sec:wcdm}

The dangers of confusing the signatures of modified geometry and modified gravity have previously been highlighted in the literature \citep[see for example][]{simpsonp09}. In order to claim that we are testing GR and not simply \lcdm  we must therefore allow greater freedom in the cosmic expansion history $H(z)$. This is inevitably accompanied by some loss of precision, so we shall restrict ourselves to adding one extra parameter. In Figure \ref{fig:curvature} we utilise both $H_0$ and BAO, in combination with CFHTLenS, RSD and the small angular scales of WMAP7, \red{conservatively omitting the ISW information for the reasons outlined in Section~\ref{sec:isw}}. While the green contour in the left panel corresponds to the standard flat \lcdm background, the blue contours presented in the left and centre panels allow the effective dark energy equation of state $w$ to deviate from $-1$. The centre panel illustrates a degeneracy which causes the confidence contours to broaden. More positive values of $w$ lead to a greater prevalence of dark energy at higher redshift (for a given value of $\om$) and thus a slowing of the growth of structure. This can be compensated by enhancing the value of $\mu$, which accelerates the growth of structure, leading to a similar growth history $f(z)$ to GR.  For a \wcdm cosmology the mean and 68 per cent confidence intervals are given by $\mu_0 = -0.59 \pm 0.34$, $\Sigma_0 = -0.19 \pm 0.11$, and $w = -1.19 \pm 0.1$. More detailed results are presented in Table \ref{tab:results}.

\subsection{$o$\wcdm}

\begin{figure*}
\includegraphics[width=175mm]{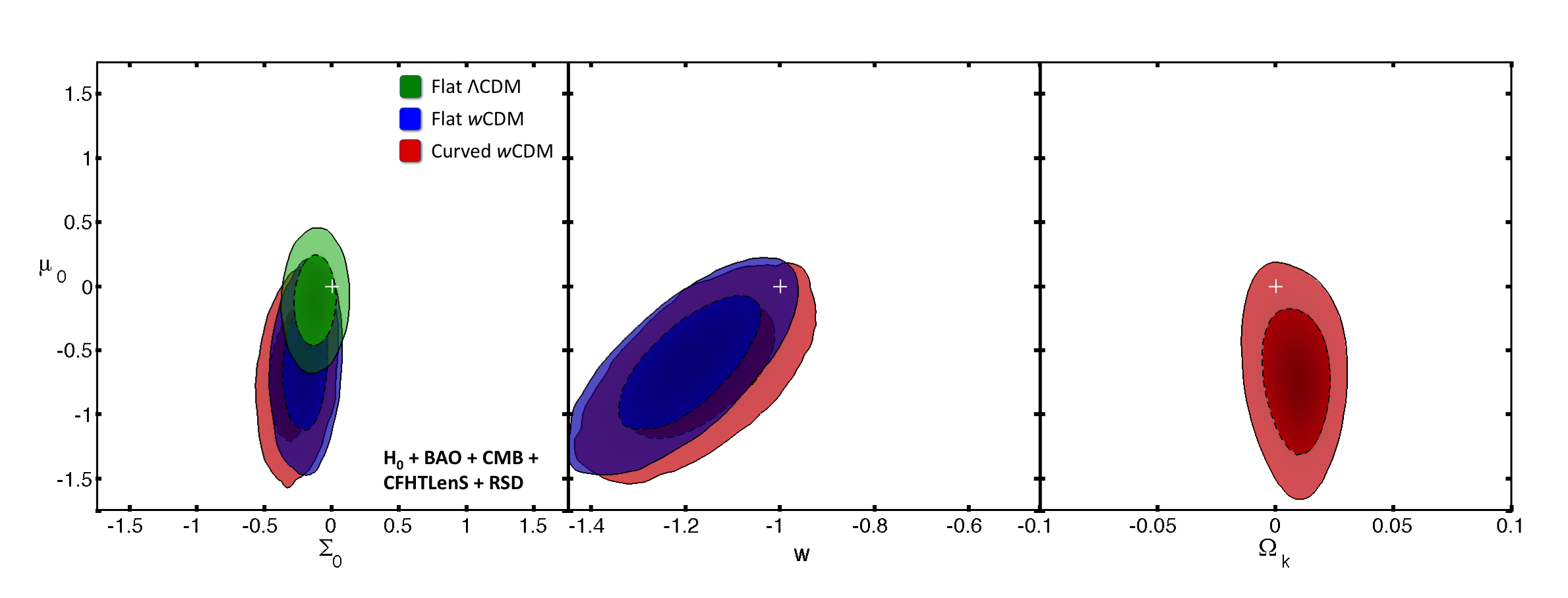}
\caption{ \label{fig:curvature}  In this figure we extend our analysis to more general expansion histories. All of the above contours use a combination of cosmic shear and redshift space distortions, combined with the geometric constraints from $H_0$, BAO, and WMAP7 ($\ell \geq 100$).  The green set of contours in the left panel corresponds to the background expansion of flat \lcdm. For the blue contours in the left and centre panels we allow the effective dark energy equation of state to deviate from $w=-1$, highlighting a significant degeneracy in the $w-\mu_0$ plane. Finally,  the red contours appearing in each of the panels allow non-zero curvature in addition to the \wcdm background.} 
\end{figure*}

We further generalise the expansion history by allowing for non-zero values of the global curvature $\Omega_k$, using the same combination of data sets as in \S \ref{sec:wcdm}. The red contours in all three panels of Figure  \ref{fig:curvature} demonstrate the relatively small impact that non-flat geometries have on our confidence limits. The mean and 68 per cent confidence intervals are given by \blue{$\mu_0 = -0.65 \pm 0.34$, $\Sigma_0 = -0.26 \pm 0.12$, $\Omega_k = 0.096 \pm 0.087$, and $w = -1.16 \pm 0.10$}.

\section{Alternative Parameterisations} 
\label{sec:alt_params}

In this section we consider the relationship between the $\left( \mu, \Sigma \right)$ parameterisation adopted in this work, and three other popular parameterisations. 

\subsection{$\gamma$ Parameter}

A widely adopted parameterisation for quantifying anomalous structure growth is the growth index $\gamma$ \citep{1998ApJ...508..483W, 2005PhRvD..72d3529L}

\[ \label{eq:gamma_f}
f(z) \simeq \om^{\gamma}(z) \, ,
\]

\noindent where setting $\gamma = 0.55$ offers an estimate of the growth factors in  \wcdm cosmologies with sub-percent accuracy. Equation (23) from \cite{2007APh....28..481L} gives us the relation

\[
\gamma  = \frac{3 \left[ 1 - w - \frac{\mu(a)}{1 - \om(a)} \right]}{5 - 6 w} \, ,
\]

\noindent for a constant equation of state $w$. In a flat \lcdm Universe, this parameter may be expressed as a simple function of $\mu_0$ and $\ov$,

\begin{figure}
\includegraphics[width=80mm]{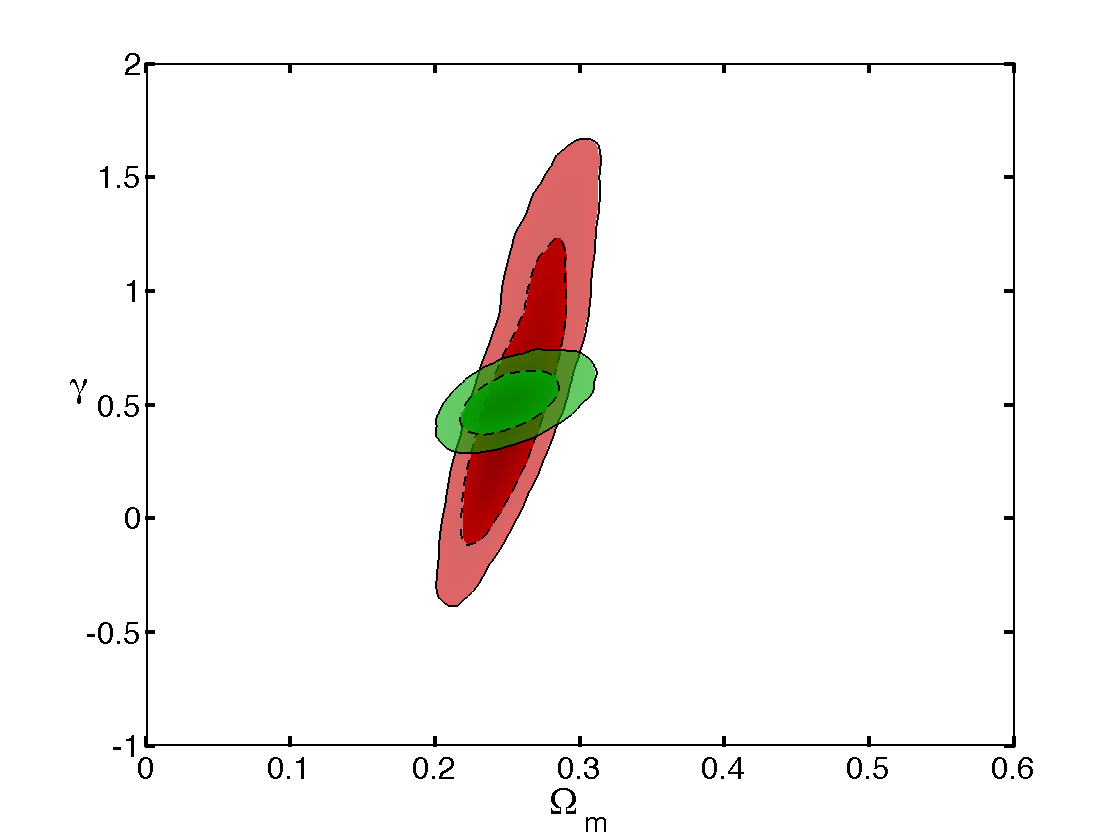}
\caption{  Confidence limits in the $\gamma$ - $\om$ plane, as derived from redshift space distortions (WiggleZ + 6dFGS) in green, and cosmic shear (CFHTLenS) in red (68 and 95 per cent CL), using small-angle anisotropies from WMAP7 ($\ell \geq 100$), and a prior on $H_0$ as auxillary data sets. Note that here $\gamma$ is the only modified gravity parameter, as we have set $\Sigma(a)=0$. \label{fig:EGgamma}}
\end{figure}

\[ \label{eq:gamma_mu}
\gamma  = \frac{6}{11} \left[1 - \frac{\mu_0}{2 \ov} \right]  \, .
\]

\begin{figure}
\includegraphics[width=80mm]{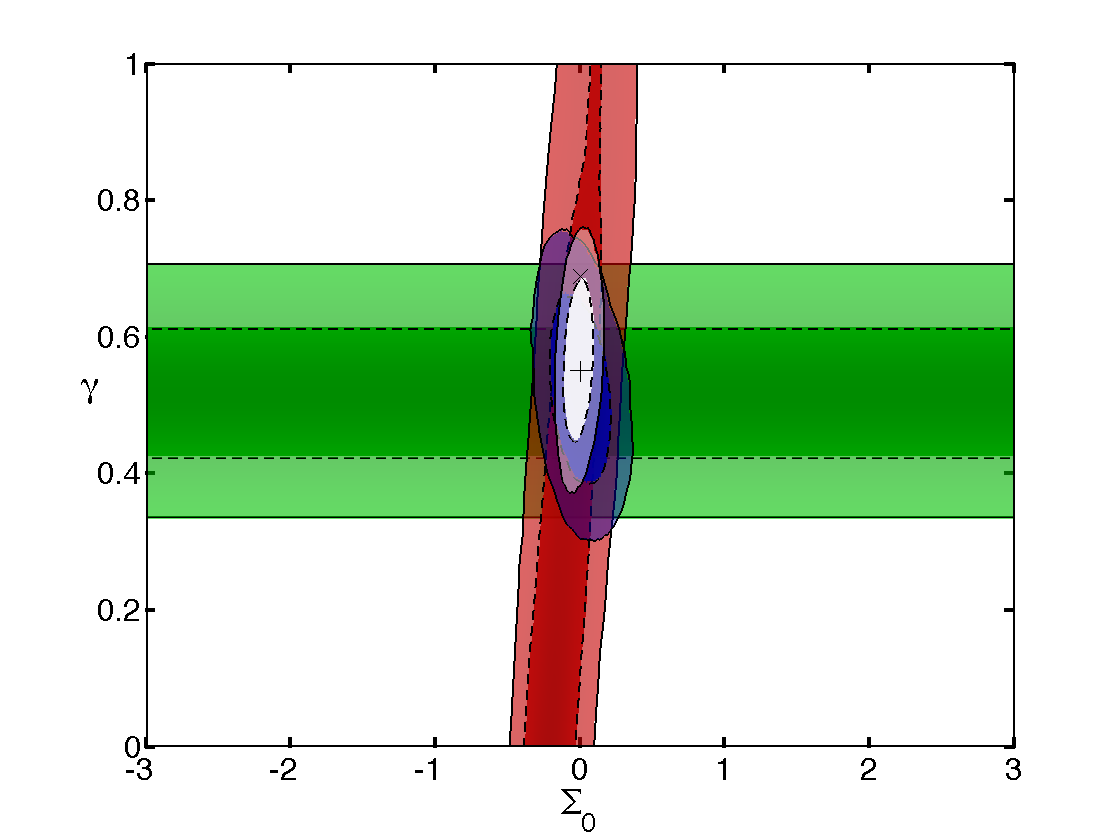}
\caption{ The same combinations of data as the left hand panel of Figure \ref{fig1hb}, except we have replaced the $\mu_0$ parameter with $\gamma$ using the relation given by equation (\ref{eq:gamma_mu}). \label{fig1h_gamma}}
\end{figure}

Applying the prescription for the growth rate in equation (\ref{eq:gamma_f}) with a constant value the $\gamma$ parameter does not offer a perfect reproduction of the growth of density perturbations in GR.  This relationship provides a useful means to derive constraints in terms of $\gamma$ without compromising the precision of the calculation. Constraints on this quantity are presented in the $\gamma-\om$ plane in Figure \ref{fig:EGgamma}. Gravitational lensing is not uniquely defined when using the $\gamma$ parameter alone, so here we have simply set $\Sigma_0 = 0$ such that the lensing potential is unaltered from its conventional form. We find a value of $\gamma = 0.52 \pm 0.09$ for the case of a \lcdm background.  \red{In Figure \ref{fig1h_gamma}, we then let $\Sigma_0$ vary, presenting constraints in the $\gamma-\Sigma_0$ plane.}

\subsection{$\EG$ Parameter}

\begin{figure}
\includegraphics[width=80mm]{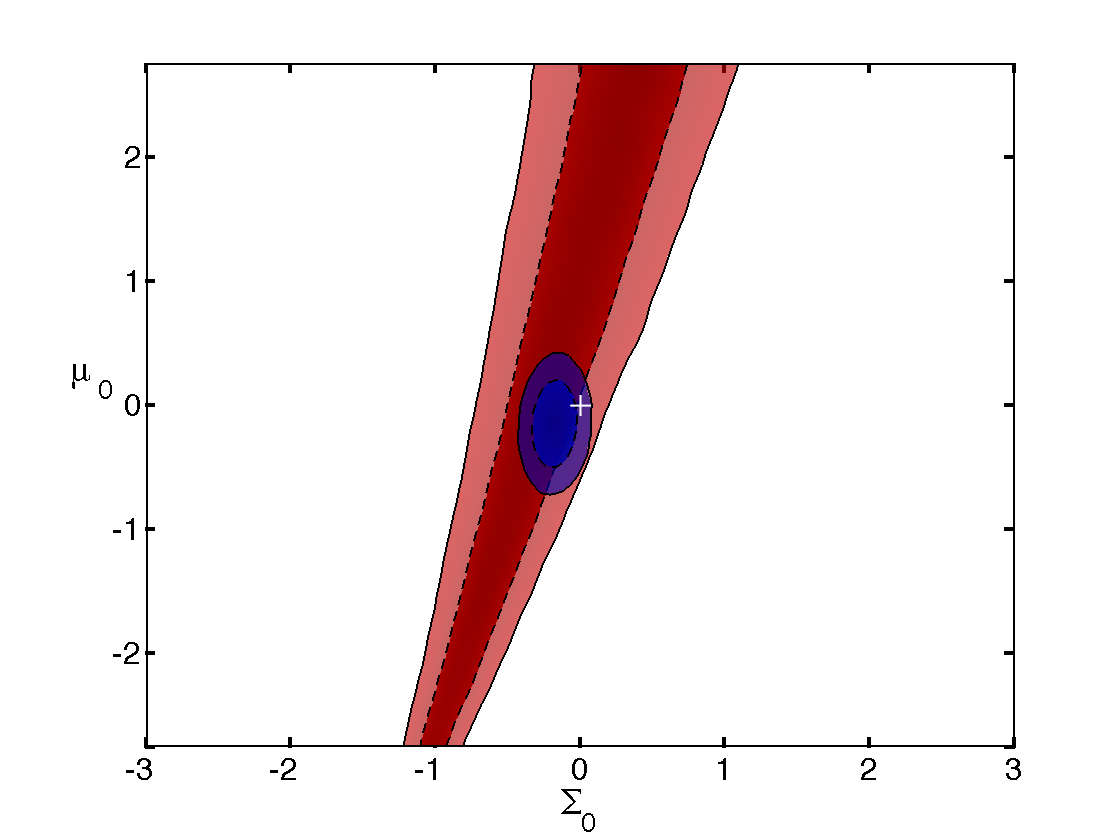}
\caption{ \label{fig:EG} Confidence limits on the $\Sigma_0$ and $\mu_0$ parameters derived from the measurement $\EG(z=0.32) = 0.392 \pm 0.065$ \citep{Reyes2010}. The inner blue contours are the same as those in Figure \ref{fig1hb}. Here we assume a \lcdm cosmology. }
\end{figure}

Proposed by \citet{Zhang2007},  the $\EG$ parameter encompasses contributions from both gravitational lensing and the motions of galaxies. This is in the form of a ratio which by construction nullifies contributions from both linear galaxy bias and the amplitude of the matter power spectrum. The result is a quantity which is sensitive to both the lensing potential and the rate of structure formation

\[ \label{eq:EG_theory}
\EG \equiv \frac{\nabla^2 (\Phi + \Psi)}{3 H_0^2 (1+z) \beta \delta}  \, ,
\]

\noindent where $\beta \equiv f/b$. This technique has been applied to SDSS data in \citet{Reyes2010}. We cannot directly measure $\EG$ with the CFHTLenS and WiggleZ datasets, since they do not overlap in sky coverage. However we can perform a direct mapping of the result from \cite{Reyes2010}  onto the $\left( \mu, \Sigma \right)$ plane, utilising the relationship

\[ \label{eq:EG}
\EG \simeq \frac{\om \left[ 1 + \Sigma(\bar{z}) \right]}{f(\bar{z})} \, ,
\]

\noindent where the dependence on $\mu$ arises through its influence on $f(\bar{z})$, the linear growth rate at the mean redshift of the survey. \red{The results are presented in Figure~\ref{fig:EG}, where the red contours correspond to the confidence limits, in the $\mu_0 - \Sigma_0$ plane, }arising from a combination of the measurement $\EG(z=0.32) = 0.392 \pm 0.065$ \citep{Reyes2010} with the BAO data from Section \ref{sec:BAO} which provide a constraint on $\om$. The CMB does not provide further assistance in this case because the $\EG$ parameter is insensitive to the amplitude of perturbations. For comparison, the inner blue contours match those in the right hand panel of Figure \ref{fig1hb}.
%CH re-order here so the Fig comes first and the caveats second
We note that GR can readily generate departures from the conventional $\EG$ value. For example in equation (\ref{eq:EG}) the ratio $\om / f(\bar{z})$ changes when $\Omega_k \neq 0$ or $w \neq -1$. Similarly, radically different modified gravity models can generate an $\EG$ value which appears consistent with the \lcdm value. This is apparent from the degeneracy contour seen in Figure \ref{fig:EG}, where large values of both $\mu$ and $\Sigma$ enhance  the growth rate and lensing signal, leaving their ratio in equation (\ref{eq:EG_theory}) unchanged.

\subsection{Gravitational Slip}

\begin{figure}
\includegraphics[width=80mm]{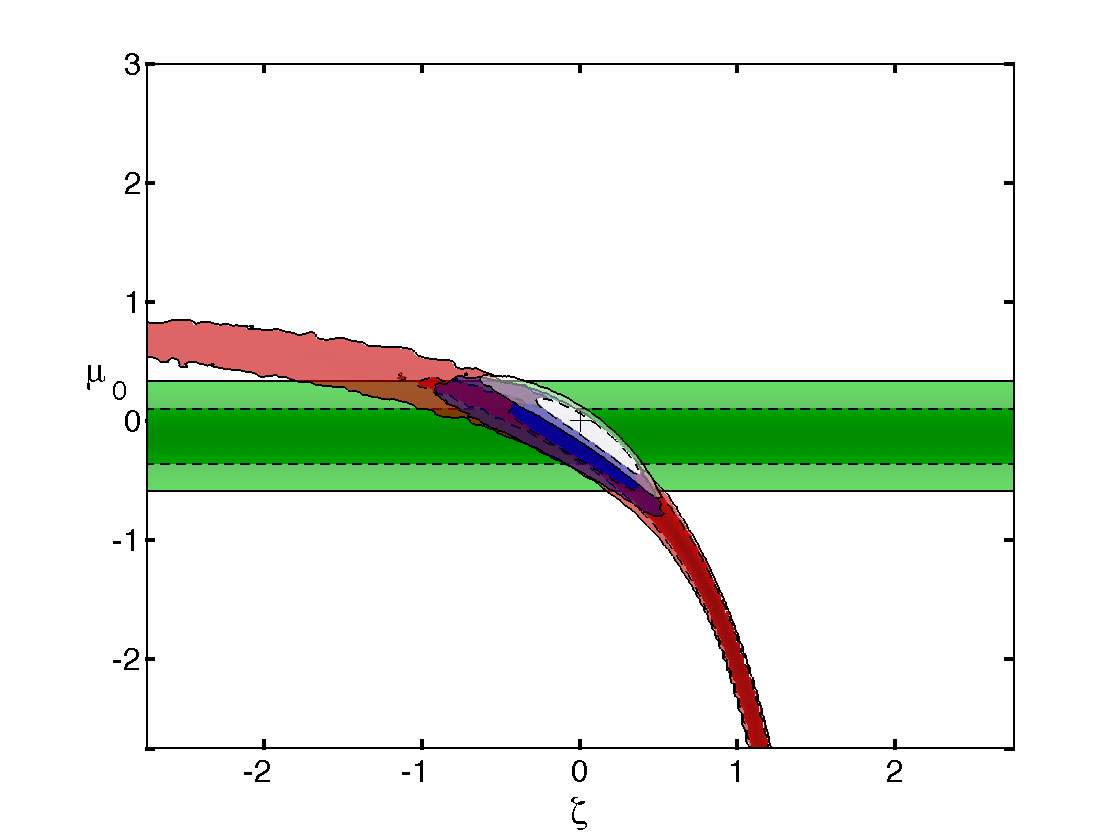}
\caption{ The same combinations of data as the left hand panel of Figure \ref{fig1hb}, except here we have replaced the $\Sigma_0$ parameter with the gravitational slip $\zeta_0$ as defined in equation (\ref{eq:slip}). \label{fig1h_slip}}
\end{figure}

Rather than modulating the sum of the two potentials using $\Sigma(a)$, it is often convenient to work in terms of the ratio of the potentials. One definition in this form is known as the gravitational slip $\zeta$,

\[ \label{eq:slip}
\zeta \equiv 1 - \frac{\Psi}{\Phi} \, ,
\]

\noindent which we expect to be zero in GR, in the absence of anisotropic stress. This has been the focus of much theoretical exploration, for example \citet{2010PhRvD..81j4020F} demonstrate that for second-order theories we generally expect to find this value scales with the Newtonian potential and its time derivative. 

Recasting $\zeta$ in terms of \musigma using equations (\ref{eq:mod_psi}) and (\ref{eq:mod_psiphi})

\[
\zeta(a) = \frac{2 \left[ \Sigma(a) - \mu(a) \right]}{1 + 2 \Sigma(a) - \mu(a)}  \, .
\]

\noindent As an illustration in Figure \ref{fig1h_slip} we reformulate the x-axis from Figure \ref{fig1hb} in terms of $\zeta_0$, the gravitational slip evaluated at $z=0$. One striking feature of these contours are the long tails out to negative values of $\zeta$ and $\mu$, corresponding to where the denominator $\Phi$ ventures close to zero. This suggests that the slip parameter may possess a pathological likelihood surface, and so for the purposes of performing Monte Carlo evaluations these contours are more safely generated as a derived parameter.

\subsection{Gravitational Potentials}
\begin{figure}
\includegraphics[width=92mm]{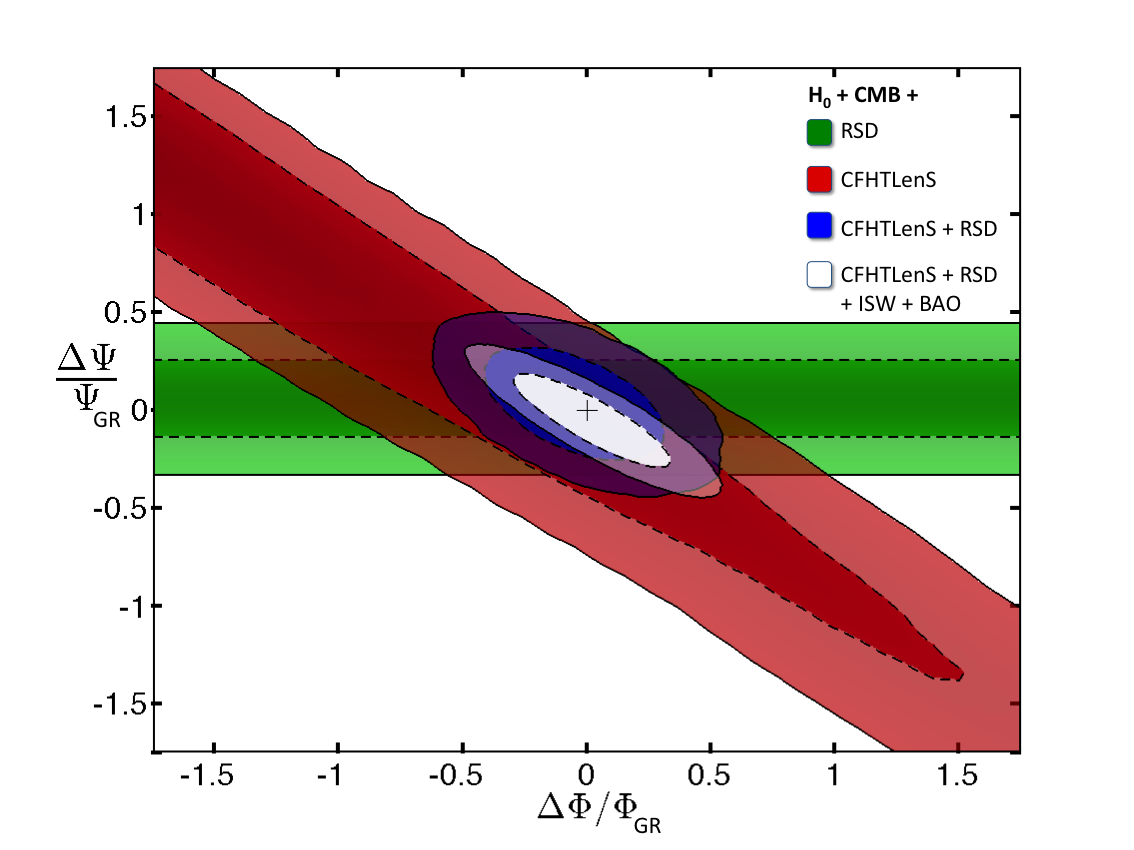}
\caption{ Here we explore fractional deviations in the two gravitational potentials, the Newtonian potential $\Psi$ and the curvature potential $\Phi$, from the GR value at $z=0.5$. The prescription for this is given by equation (\ref{eq:psiphi}).  The contours represent the same combinations of data as those in the left hand panel of Figure \ref{fig1hb}.  \label{fig:phi}}
\end{figure}

\red{The redshift-space distortion data, used in this analysis, exhibits a peak sensitivity to a modified gravity scenario at approximately $z \simeq 0.5$ (see Section~\ref{sec:theory}). In Figure \ref{fig:phi} we therefore consider fractional deviations in the two gravitational potentials at this redshift, relative to their GR prediction for the same mass distribution. } These are given by

\ba 
\frac{\Delta \Psi}{\Psi} &=& \mu(z) \, ,\cr  
\frac{\Delta \Phi}{\Phi}  &=& 2 \Sigma(z) - \mu(z) \, . \label{eq:psiphi}
\ea

\noindent where $\Delta \Psi \equiv \Psi - \PsiGR$.   \red{For our conservative combined lensing, RSD, $H_0$ and high $\ell$ WMAP7 data analysis we find cosmological deviations in the Newtonian potential and curvature potential from the prediction of General Relativity to be $\Delta \Psi/\Psi = 0.05 \pm 0.25$ and  $\Delta \Phi/\Phi = -0.05 \pm 0.3$  respectively ($68$ per cent CL).}

\section{Theoretical Models} 
\label{sec:theory}
Below we briefly review some of the theoretical models which could generate a departure from $\mu_0 = \Sigma_0 = 0$, and interpret the implications of our results.  There is such a plethora of modified gravity models, that no single choice of parameterisation can adequately encompass all of them. This is a situation reminiscent of the dark energy equation of state, $w(z)$, except here we are faced with uncertainty not only in the functional form of the time-dependence, but also in its scale-dependence.  So how can we relate a given \musigma constraint to a specific model? The observed parameters $\hat{\mu}_0$ and $\hat{\Sigma}_0$ may be interpreted as a weighted integral over the true functional form $\mu(k, z)$, such that
%  \blue{ Generalising the method presented in \citet{2003MNRAS.343..533D}, the} 
\[ \label{eq:wt_fn}
\hat{\mu}_0 = \int \! \! \! \int \phi (k, z) \, \mu (k, z) \frac{\ov}{\ov (z)} \, \ud k \, \ud z  \, .
\]

\noindent \blue{If we perform a scale- and time-dependent principal component analysis \citep[see for example][]{ZhaoPCA}}, 
then the weight function $\phi(k, z)$ may be \blue{expressed in terms of} the principal components $e_i(k,z)$ and the errors associated with their corresponding eigenvalues $\sigma(\alpha_j)$ \citep{2006PhRvD..73h3001S}, such that

\[ \label{eq:pca}
\phi(k, z) = \frac{\sum_i e_i(k, z) \int \! \! \! \int e_i (k', z') \ud k' \, \ud z'/ \sigma^2(\alpha_i)}{\sum_j \left[ \int \! \! \! \int e_j(k'', z'') \ud k'' \, \ud z''\right]^2/ \sigma^2(\alpha_j)}  \, .
\]

\begin{figure}
\includegraphics[width=80mm]{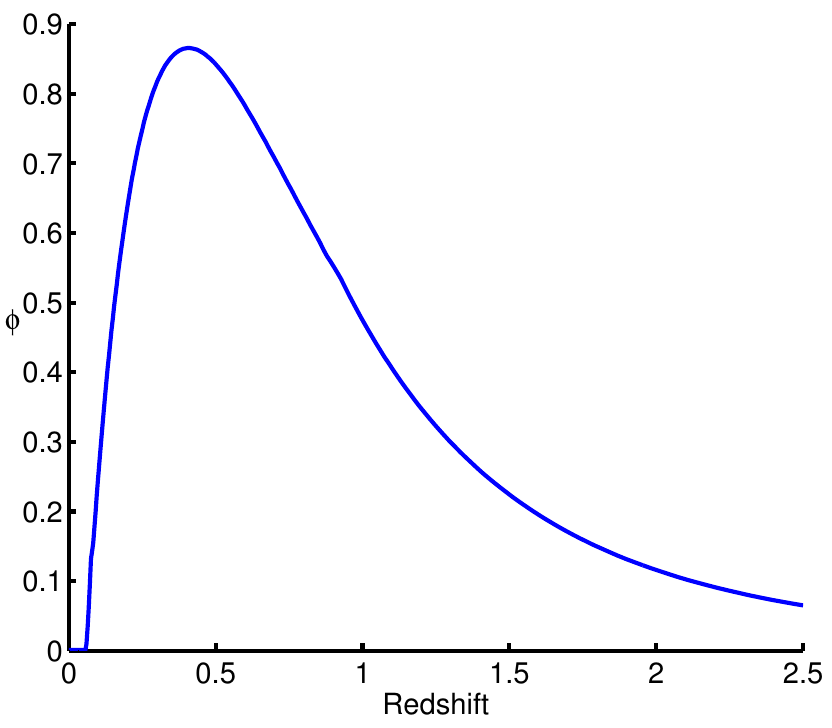}
\caption{ The redshift sensitivity \blue{of the modified gravity parameter $\mu_0$}. The weight function $\phi(z)$ is defined in equation (\ref{eq:wt_fn}) \blue{and evaluated with equation (\ref{eq:pca})}. \label{fig:weight_function}}
\end{figure}

\noindent  The analysis of redshift space distortions in \citet{Blake2012MNRAS} includes information from the galaxy power spectrum up to a maximum wavenumber $k_{\rm{max}} = 0. 2 \hmpc$, corresponding to the regime over which the density and velocity fields are sufficiently linear for our theoretical models to remain valid. Since the number of Fourier modes increases towards higher $k$, the scale-dependent component of $\phi(k, z)$ peaks close to this value of $k_{\rm{max}}$\blue{, and $\phi(k, z)=0$ for $k>k_{\rm{max}}$}. We evaluate the redshift-dependence of the weight function $\phi(z)$ associated with the combined WiggleZ and 6dFGS data of Figure \ref{fig:fsig8_data}, following the prescription of \citet{2006PhRvD..73h3001S}, and this is shown to peak at $z \sim 0.5$ as illustrated in Figure \ref{fig:weight_function}.  In the following subsections we utilise the weight function $\phi(z)$ presented in Figure \ref{fig:weight_function} to map specific examples of theoretical models onto our parameter space, by evaluating equation (\ref{eq:wt_fn}). However as stressed earlier, we do not aim to place rigorous parameter constraints on any particular family of models. 

%NB \Phi_gamma in \cite{simp09} is identical to \Phi_\mu0 due to the relation (\ref{eq:gamma_mu})

\subsection{$f(R)$}

A more general form of the Einstein-Hilbert action replaces the Ricci scalar R with an arbitrary function $f(R)$ such that

\[
S = \int f(R) \sqrt{-g} \, \ud^4 x  \, ,
\]

\noindent where $g$ is the determinant of the metric tensor. This defines the broad class of $f(R)$ models. One of the most difficult tasks for any modified gravity model attempting to replace dark energy is to satisfy the stringent Solar System constraints, and most natural choices of the function $f(R)$ fail to do so. The subset of $f(R)$ models which have attracted interest are those which employ the so-called chameleon mechanism, where departures from GR are strongly suppressed in regions where $R$ is large, only emerging when $R$ is sufficiently small. Our location within the potential well of the Sun and the Milky Way halo may be sufficient to shield us from this unusual gravitational behaviour. 

For a particular subset of $f(R)$ models which are capable of satisfying Solar System tests, the departure from GR may be characterised as \citep{Zhao2011} 

\[
\mu(k, a) = \frac{1}{3 + 3(aM/k)^2} \, ,
\]

\noindent where the scalaron mass $M(a)=1/\sqrt{3 \, \ud^2 \! f / \ud R^2}$. For any given redshift and wavenumber, the value of $\mu$ lies in the range $0 \leq \mu < \frac{1}{3}$. This generically enhances growth, so we expect this family of models to lie vertically above the point $(0,0)$ in Figure \ref{fig1hb}. We parameterise $M=M_0 a^{-\sigma}$ and take as an example $M_0 = 0.02 \hmpc$  and $\sigma=3$, \blue{corresponding to the type of model explored in \citet{Zhao2011}}.  In $f(R)$ models the lensing potential for a given mass distribution is unchanged from the case of GR, and so $\Sigma_0^{f(R)} = 0$.  Our measure of $\mu$ is dominated by the RSD data and by evaluating equation (\ref{eq:wt_fn}) this model ought to map onto our \musigma plane at a value of $\mu_0^{f(R)} = 0.81$. This is despite $\mu(a)$ never rising above $1/3$, but the stronger deviation from GR at high redshift compared to our parameterisation allows a higher effective value of $\mu_0$ to be generated. Our data therefore disfavour the more aggressive $f(R)$ models\blue{, in agreement with the findings of \citet{Zhao2011}}.

\subsection{DGP}

The higher-dimensional model proposed by \citet*{2000PhLB..485..208D} has proved one of the most popular gravitational models for theoretical and experimental exploration.  It is of particular interest due to its physical motivation and specific observational prediction, namely a value of the growth index  $\gamma = \frac{11}{16}$, or equivalently $\mu_0 = -\frac{25}{48} \Omega_\Lambda; \Sigma_0 = 0$. However in its most natural form there appears significant disparity with observational data \citep[see for example][]{2009PhRvD..80f3536L}. This can be bypassed by introducing a cosmological constant, although this somewhat compromises the motivation of the DGP model. 

In general the behaviour of DGP is a scale-independent suppression of the growth of structure. Like $f(R)$, the lensing potential is unaltered so $\Sigma_0^{\rm{DGP}}=0$. Adopting a fiducial value $\Omega_\Lambda = 0.27$ leaves us with $\mu_0^{\rm{DGP}} = -0.38$. This negative value of $\mu$ remains consistent with our data.

\subsection{Newtonian Limit}

The point at $\mu_0=0; \Sigma_0 = -0.5$ corresponds to a scenario where the present-day cosmological gravity resembles Newtonian gravity. That is, the angular deflection of light is half the value predicted by Einstein at $z=0$, while the forces experienced by galaxies remain unchanged. This model was famously rejected in favour of General Relativity by \citet{1920RSPTA.220..291D}. The difference here is that we are performing this test on cosmological scales. Whether we combine our gravitational data with either BAO or the $H_0$ prior, the results are the same, a value of $\Sigma_0 = - 0.5$ is strongly disfavoured by the data.
%We denote this benchmark with a circle in Figure \ref{fig1hb}. 

Another regime of interest is that of $\mu_0<-1$, for this corresponds to the anti-gravitational behaviour of galaxies, in that they are repulsed from clusters and attracted towards voids. One might naively expect that if dark energy is gravitational in origin, then the repulsive nature of the expansion may also be seen within the perturbations. This region is disfavoured by the data when assuming a flat \lcdm background, but this is no longer the case for more general backgrounds since the regime $w<-1$ permits more negative values of $\mu_0$, as seen in the central panel of Figure \ref{fig:curvature}. 

\subsection{Interacting Dark Energy}

When conducting cosmological tests of gravity, in this work and in others \citep{Daniel2010, Bean2010, Zhao2010, 2010arXiv1011.2106S,  Reyes2010, 2011A&A...530A..68T, ZuntzISW, Zhao2011,  Rapetti2012}, we necessarily make certain assumptions such as the lack of anisotropic stress. One assumption not often made explicit is that dark energy exhibits negligible interactions with dark matter, aside from their mutual gravitational repulsion. If the two fluids were to exchange energy and momentum, this readily alters the growth of structure \citep{2000PhRvD..62d3511A, sjp, simpscat}, and so could generate a false signature of modified gravity. 

In the simplest interacting models, we would expect to see an alteration in the growth of large scale structure, but no direct change to the lensing potential. Thus detecting a non-zero value of $\Sigma_0$ is a prerequisite for confirming a truly modified gravity scenario. That is to say, we cannot perform a cosmological test of gravity without gravitational lensing. 

As an illustrative example, we consider the scenario of dark energy decaying into a form of dark matter. If this results in the dark matter density following a power law $\rho = \rho_0 a^{-3+\epsilon}$, the growth rate of cosmic structure is subject to a constant decrement \citep{sjp}

\[
f(a) = \om^\gamma(a) - \frac{67}{55} \epsilon \, .
\]

\noindent Large values of $\epsilon$ lead to a significant quantity of dark energy in the early Universe, and as such are disfavoured by observations of the CMB. Taking $\epsilon=0.01$, which is small enough not to perturb the primary anisotropies of the CMB, leads to an effective value $\mu_0^{\epsilon} =-0.99$ and $\Sigma_0^{\epsilon}  = 0$, \red{a region which is disfavored by our results.}

\noindent

\section{Conclusions} 
\label{sec:conclusions}
In this work we find no indication of a departure from General Relativity on cosmological scales, having explored the behaviour of non-relativistic and relativistic particles in the WiggleZ, 6dFGS, and CFHTLenS data. Both the motions of galaxies and the deflection of their emitted light conform to the predictions of Einstein's theory, at the level of precision permitted by current data. We place limits on our gravitational parameters  $\mu_0 = 0.05 \pm 0.25$ and $\Sigma_0 = 0.00 \pm 0.14$ for a \blue{flat} \lcdm expansion history. This corresponds to deviations in the present-day time dilation and spatial curvature of $\Delta \Psi/\PsiGR = 0.05 \pm 0.25$ and $\Delta \Phi / \PhiGR = -0.05 \pm 0.3$ respectively. \blue{However these errors are significantly correlated, as shown in Figure \ref{fig:phi}.}

For ground-based and solar system tests, where light-travel time is short, it is relatively straightforward to conduct a geometric measurement. Cosmological experiments do not share this luxury, and considerable effort is required in order to determine the spatial positioning of the test subjects. The precision at which we map the geometry of the Universe remains one of the limiting factors in determining its gravitational potentials. This is particularly apparent when increased freedom in the cosmic expansion history is allowed, arising from a variable dark energy equation of state or non-zero global curvature. Further geometric data could be included in the analysis, such as those derived from the light curves of Type Ia supernovae. However combining a large number of datasets leaves us increasingly exposed to the effects of systematic errors. For the purposes of gravitational lensing, another source of uncertainty is the form of the matter power spectrum on nonlinear scales, and this is an issue which will likely become more acute with the advent of surveys spanning a large fraction of the sky. Higher resolution N-body simulations can help to a certain extent, but modelling the influence of baryons may prove particularly problematic.

The absence of a compelling theoretical rival to GR makes a suitable choice of gravitational parameters rather uncertain. In order to minimise the number of degrees of freedom, and to maintain sensitivity to a broad range of models, we adopted a scale-independent modification to the Newtonian and curvature potentials. The scale-dependence of our results is implicit in the wide range of sensitivities each probe possesses. The cosmic shear signal from CFHTLenS is sourced by density perturbations of the order ${\sim}5 \hinvmpc$, while the redshift space distortions from WiggleZ involve Fourier modes with wavelengths over $30 \hinvmpc$. Furthermore the ISW effect causes the largest angular scales of the CMB to receive significant contributions from gravitational perturbations in excess of ${\sim} 500 \hinvmpc$. If any inconsistency were seen between these results this could have been an indication of scale-dependent behaviour. 
Our reasons for focussing on a time dependent variation that mimics the dominance of dark energy rather than using constant  $\mu$ and $\Sigma$ values are twofold. First our motivation for seeking modification to General Relativity stems from dark energy, a phenomenon which only appears at late times. Furthermore, if we were to investigate constant $\mu$ and $\Sigma$ parameters, our constraints would be dominated by the primary anisotropies of the cosmic microwave background, which as shown in previous studies \citep{Bean2010, ZuntzISW} is consistent with a GR framework.  

Solar system experiments have already eliminated many potential modifications to GR which could have provided a natural explanation for the accelerating Universe. Here we have begun to disfavour some of those which remain. While the volume of parameter space available to modified gravity models continues to shrink, the cosmological constant retains its position as the most viable solution to the dark energy problem.

\section{Acknowledgments}
We would like to thank Gong-Bo Zhao, Pedro Ferreira, John Peacock, Benjamin Joachimi, and the referee Joe Zuntz for helpful discussions.

This work is based on observations obtained with MegaPrime/MegaCam, a joint project of CFHT and CEA/DAPNIA, at the Canada-France-Hawaii Telescope (CFHT) which is operated by the National Research Council (NRC) of Canada, the Institut National des Sciences de l'Univers of the Centre National de la Recherche Scientifique (CNRS) of France, and the University of Hawaii. This research used the facilities of the Canadian Astronomy Data Centre operated by the National Research Council of Canada with the support of the Canadian Space Agency.  We thank the CFHT staff for successfully conducting the CFHTLS observations and in particular Jean-Charles Cuillandre and Eugene Magnier for the continuous improvement of the instrument calibration and the {\sc Elixir} detrended data that we used. We also thank TERAPIX for the quality assessment and validation of individual exposures during the CFHTLS data acquisition period, and Emmanuel Bertin for developing some of the software used in this study. CFHTLenS data processing was made possible thanks to significant computing support from the NSERC Research Tools and Instruments grant program, and to HPC specialist Ovidiu Toader.  The N-body simulations used in this analysis were performed on the TCS supercomputer at the SciNet HPC Consortium. SciNet is funded by: the Canada Foundation for Innovation under the auspices of Compute Canada; the Government of Ontario; Ontario Research Fund - Research Excellence; and the University of Toronto.  The early stages of the CFHTLenS project were made possible thanks to the support of the European CommissionÕs Marie Curie Research Training Network DUEL (MRTN-CT-2006-036133) which directly supported members of the CFHTLenS team (LF, HH, BR, MV) between 2007 and 2011 in addition to providing travel support and expenses for team meetings.

FS, CH, H. Hoekstra \& BR acknowledge support from the European Research Council under the EC FP7 grant numbers 240185 (FS, CH), 279396 (HH) \& 240672 (BR). CB acknowledges the support of the Australian Research Council through the award of a Future Fellowship. LVW acknowledges support from the Natural Sciences and Engineering Research Council of Canada (NSERC) and the Canadian Institute for Advanced Research (CIfAR, Cosmology and Gravity program).  TE is supported by the Deutsche Forschungsgemeinschaft through project ER 327/3-1 and the Transregional Collaborative Research Centre TR 33 `The Dark Universe'. H. Hildebrandt is supported by the Marie Curie IOF 252760 and by a CITA National Fellowship. H. Hoekstra also acknowledges support from Marie Curie IRG grant 230924 and the Netherlands Organisation for ScientiÞc Research grant number 639.042.814.   TDK acknowledges support from a Royal Society University Research Fellowship.  YM acknowledges support from CNRS/INSU (Institut National des Sciences de l'Univers) and the Programme National Galaxies et Cosmologie (PNCG).  LF acknowledges support from NSFC grants 11103012 and 10878003, Innovation Program 12ZZ134 and Chen Guang project 10CG46 of SMEC, and STCSM grant 11290706600.  MJH acknowledges support from the Natural Sciences and Engineering Research Council of Canada (NSERC).  TS acknowledges support from NSF through grant AST-0444059-001, SAO through grant GO0-11147A, and NWO. MV acknowledges support from the Netherlands Organization for Scientific Research (NWO) and from the Beecroft Institute for Particle Astrophysics and Cosmology.  

{\small Author Contributions: All authors contributed to the development and writing of this paper.  The authorship list reflects the lead authors of this paper (FS, CH, DP, CB, MK) followed by two alphabetical groups.  The first alphabetical group includes key contributers to the science analysis and interpretation in this paper, the founding core team and those whose long-term significant effort produced the final CFHTLenS data product.  The second group covers members of the CFHTLenS team who made a significant contribution to either the project, this paper or both.  CH and LVW co-led the CFHTLenS collaboration.}

\appendix

\section{Treatment of the CMB} 
\label{sec:appendix}

In our analysis we use the $\ell \geq 100$ WMAP7 data from the $C_\ell^{TT}$ and $C_\ell^{TE}$ power spectra to constrain geometric cosmological parameters, as well as a benchmark for the amplitude of matter perturbations. In contrast, the $\ell < 100$ multipoles provide additional information on the gravitational potentials through the ISW. Discarding the low multipoles would severely compromise our ability to constrain the primordial amplitude of density perturbations. This is because on small angular scales there is a strong degeneracy between the optical depth $\tau$ and the amplitude $A_s$, since both parameters induce a simple scaling of the amplitude.  To circumvent this problem we first run a likelihood analysis on the low multipoles $2 \leq \ell < 100$, marginalising over our full parameter set including $\mu_0$ and $\Sigma_0$. This establishes a prior on $\tau$, which we then use in our analysis of the data, \blue{except in those cases where the low multipoles are included}. Despite the addition of modified gravity parameters, we find little degradation in the constraint on $\tau$ compared to the standard WMAP7 result which assumes GR. This is due to the unique signature left by $\tau$ at the large scales of the $C_{\ell}^{TE}$ spectrum, which cannot be replicated by the other parameters. 

\begin{figure}
\includegraphics[width=80mm]{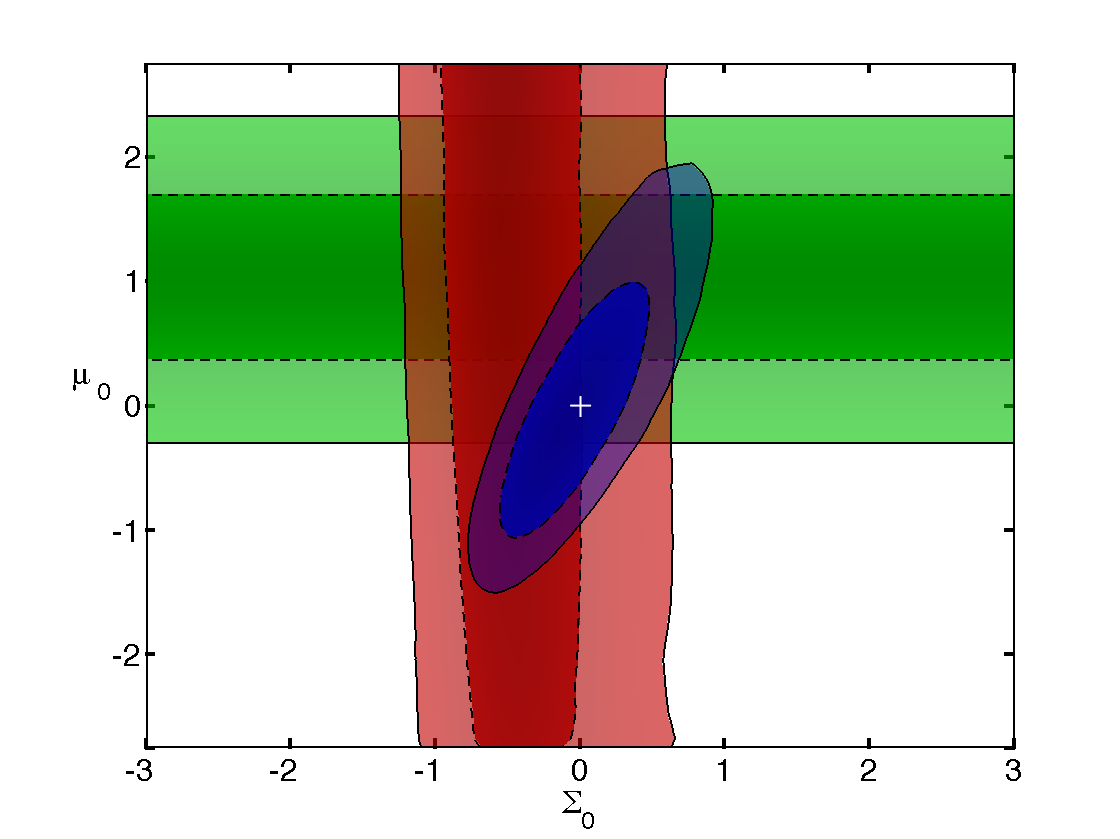}
 \caption{Constraints on our modified gravity parameters $\mu_0$ and $\Sigma_0$ from redshift space distortions (green), weak lensing (red), and combined (blue), for a flat \lcdm background.  A prior on $H_0$ is included, but there is no additional information from the CMB. The dashed contours represent the $68$ and $95$ per cent confidence intervals, while the cross denotes the prediction of GR. \label{fig1noCMB}} 
\end{figure}

To illustrate the impact of removing the WMAP7 data, we present in Figure \ref{fig1noCMB} the results for CFHTLenS $+ H_0$ in red, and WiggleZ $+ H_0$ in green.  The substantial degradation in these contours highlights the importance of including the high$-\ell$ data.

\setlength{\bibhang}{2.0em}
\setlength{\labelwidth}{0.0em}
%\bibliography{biblio}
\bibliographystyle{mn2e}
\bibliography{modgrav12}
\label{lastpage}

\end{document}